\documentclass[10pt]{article}
\begin{document}

\centerline{\Large \textbf{Qualitative evolution in higher-loop string
cosmology }}

\bigskip

\centerline{A. A. Saharian\footnote{%
E-mail address: saharyan@www.physdep.r.am}}

\bigskip

\centerline{\it Department of Physics, Yerevan State University}

\centerline{\it 1 Alex Manoogian St., 375049 Yerevan, Armenia}

\bigskip

\textbf{Abstract.} We investigate the evolution of the homogeneous and
isotropic universe within the framework of the effective string gravity with
string-loop modifications of dilaton couplings. In the case of barotropic
perfect fluid as a nongravitational source the set of cosmological equations
is presented in the form of the third-order autonomous dynamical system. The
cases are considered when this system is integrable in terms of integrals
depending on dilaton coupling functions. Without specifying these functions
we describe generic evolution of the universe, using dynamical systems
methods. The critical points and their stability are found for all regions
of the parameters. The qualitatively different phase space diagrams are
presented for the spatially-flat, closed and open universes. The case of the
tree-level models is considered separately. The issue of dilaton
stabilization is discussed within the framework of Damour-Polyakov mechanism.

\bigskip

PUCS numbers: 9880H, 9880B, 9880C

\bigskip

\section{\protect\smallskip Introduction}

The well known theorems of Hawking and Penrose \cite{Hawking} proof the
existence of spacetime singularities in a broad class of GR solutions under
certain mild conditions on the matter energy-momentum tensor. Well known
examples are black holes and the initial cosmological singularity in the
standard Big Bang scenario. It is commonly believed that in this large
curvature regions of the gravitational fields GR fails to be a completely
consistent description and should be superseded by the more fundamental
theory, in all probability, quantum in its nature. At present the most
promising candidate for the consistent quantum gravity theory and for a
unification of fundamental interactions are string theory and its recent
extensions such as M-theory \cite{green}-\cite{witten0}. In these theories
the characteristic scale in which the stringy effects become important is
Planckean size which is much smaller than the scale we can probe in recent
high energy experiments. Cosmology is only available way to test the
consequences of the theory, since the cosmological predictions originate
from physics of the early Universe, when space-time curvature may have been
of Planckean strength (about the possibilities the string scale being close
to the electroweak scale see \cite{Lykken}-\cite{Antweak}). The today
detectable remnants of primordial processes could be a test to which to
compare the theory and a number of problems of standard cosmology, like
singularity, dark matter and large scale structure, could be solved.

The standard approach to string cosmology is based on the analysis of time
dependent solutions of the effective field theory equations (see, for
example, refs \cite{binetruy}-\cite{soda}). The effective theory of gravity
resulting from string theory at low energies includes important
modifications of GR due to the presence of the additional degrees of freedom
such as dilaton, axion and etc. These fields couple to each other and to
gravity nonminimally, and can influence the dynamics significantly. The
study of cosmological consequences of these modifications has been an area
of much active research in the past. Many authors have considered various
aspects of the solutions, both in the E-frame and in the string frame. The
simplest case is the pure gravi-dilaton solution \cite{myers}, \cite{bachas}%
, \cite{mueller}, \cite{tseytlin0}, \cite{gasperini3}, \cite{kalop93}, \cite
{perry94}, which is used in \cite{gasperini3} to construct the pre-big bang
scenario of cosmological inflation. One can obtain further generalizations
of this solution for the models with spatial curvature \cite{tseytlin0}-\cite
{perry94} and with the antisymmetric tensor fields of both NS-NS and R-R
types \cite{tseytlin0}, \cite{perry94}, \cite{copeland}-\cite{ricci}, \cite
{lu}, \cite{sah2}, \cite{cop12}, \cite{lukas}, \cite{kaloper1}, \cite
{kalop00} (and references therein), either by directly solving the
corresponding cosmological equations or by using generating techniques. Such
techniques include various dualities (see, e.g., \cite{copeland}, \cite
{lids96}, \cite{meissner}-\cite{lukas}, \cite{cop13} ) and a dimensional
reduction of the higher dimensional black hole solutions \cite{behrndt}, 
\cite{poppe}. Much of the emphasis in previous work on string cosmology has
been on the problem of cosmological singularity. The attempts to address it
included the use of winding modes \cite{brand}, \cite{tseytlin2},
higher-derivative corrections \cite{madden1}, \cite{anton}-\cite{gaspven96}, 
\cite{soda}, models with generalized scalar-tensor couplings \cite{barrow1}, 
\cite{Rama}, \cite{kalop97}, inhomogeneous phase \cite{gaspmahven}, \cite
{barrow}, \cite{Giovinhom} and etc.

A large number of cosmological models have been derived in the past from a
variety of vacua of M-theory \cite{lu}, \cite{poppe}, \cite{lukas}, \cite
{kalop00}, \cite{Hawking1}, \cite{benakli}. These include solutions of type
II strings with R-R and NS-NS background fields. In particular, it has been
pointed out that cosmological solutions can be generated from p-brane
solutions by inverting the roles of the time and radial spatial coordinate.
A new way of investigating string cosmology is connected to the conjecture
due to Maldacena \cite{Maldacena}, which relates string theory in
Anti-de-Sitter spacetime to a conformal field theory living on its boundary
(see \cite{Horowitz}).

One of the basic features of superstring theories is the existence of scalar
fields, called moduli, which couple with gravitational strength and are
massless to all orders in perturbation theory. The most general modulus
field is the dilaton, whereas the other moduli parametrize the size and the
shape of the extra compact dimensions. These fields are natural partners of
the metric and play an important role in cosmology. A number of parameters
in the string effective action, such as gauge and gravitational couplings
are determined by expectation values of the moduli. As a consequence the
cosmological variations of moduli fields will lead to the corresponding
variations of the physical constants. To avoid conflict with observations
conventionally it is assumed that non-perturbative phenomena create a
potential which generate masses for the dilaton and the moduli, and at the
same time provide a mechanism of supersymmetry breaking. The favorite
mechanism of such a type is due to gaugino condensation in the gauge group
hidden sector (see, for instance, \cite{Cvet}, \cite{Quev}, and references
therein). The resulting potential for the dilaton is a combination of
exponentials and polynomials in the field. A detailed investigation of these
condensate models has demonstrated the need for at least two condensates to
form if the dilaton potential is to develop a minimum at a realistic value
(''racetrack'' models) \cite{racetrack}. Recently an alternative proposal,
relying on only one gaugino condensate, has been suggested as a method of
obtaining a minimum for the dilaton \cite{Cas}, \cite{Binstab}. In this
scenario the K\"{a}hler potential requires string desired nonperturbative
corrections. The analysis of this models indicates that it is possible to
have a minimum with zero or small positive cosmological constant \cite{Barr1}%
. The possibilities for inflation and attempts for the resolution of
cosmological moduli problem in the context of these mechanisms have been
considered in \cite{Brustinf}, \cite{Barr1}-\cite{Gailinf} (and references
therein).

Another possibility of a relaxation mechanism by which various moduli fields
are attracted towards their present vacuum expectation values due to
string-loop modifications of the dilaton couplings offers Damour-Polyakov
mechanism \cite{damour}. The idea is that nonperturbative effects associated
with higher genus corrections, may naturally generate different nonmonotonic
coupling functions of the dilaton to the other fields (about the possibility
of stabilizing the moduli fields of type II string theory or M-theory by
using the tree-level couplings to the non-trivial NS-NS and R-R form fields
see \cite{lukasstab}). Under the assumption that the different coupling
functions have extrema at some common point, the expansion of the Universe
drives the dilaton vacuum expectation value towards the extrema at which the
interaction with matter become very weak. Constraints for this mechanism
from the big bang nucleosynthesis and from the emission of massless dilatons
by the binary pulsar are studied in \cite{Vayon}. The authors of \cite
{damour1} have shown that the inflationary era with Damour-Polyakov
mechanism could solve the cosmological moduli problem and the produced
quantum fluctuations of the relevant moduli fields during this era are
naturally compatible with the observational requirements.

In \cite{saharcqg} for the general case of the dilaton coupling functions we
have derived the solutions to the string cosmology with arbitrary curvature
and with moduli and Kalb-Ramond fields as a source. These solutions were
given in terms of integrations depending on the coupling functions. For the
general case of the barotropic perfect fluid as a non-gravitational source
the set of cosmological equations can not be solved explicitly. Here we
shall present the phase-space analysis of the string cosmologies with
loop-corrections to the dilaton coupling functions and barotropic perfect
fluid. The issue of dilaton stabilization is considered within the framework
of Damour-Polyakov mechanism.

We have organized the paper as follows. In the next section the string
effective action with dilaton coupling functions of general form and its
different conformal frames are considered. Section 3 concentrates on
homogeneous and isotropic cosmological models. For the barotropic perfect
fluid the main features of the model are determined by the barotropic index
and by the function $\alpha (\phi )$. At tree-level when the dilaton
coupling functions are exponentials for the NS-NS sector fields and
constants for the R-R sector this function is a constant. In the general
case of the models with curved space the set of cosmological equations can
be presented in the form of third-order autonomous dynamical system. In
section 4 the cases are considered when this system is exactly integrable.
They include pure gravi-dilaton, stiff fluid, radiation-dominated models, as
well as two tree-level special solutions for the general barotropic index.
The section 5 presents the qualitative analysis for the spatially-flat
models, when the corresponding dynamical system is reduced to the
second-order one. For this system the critical points and their stability
are investigated. Various qualitatively different phase portraits are
presented. Another case when the corresponding dynamical system is reduced
to the second-order system are tree-level models (section 6). The spatially
curved models in the general case of dilaton coupling functions are
considered in section 7. For these models the phase space is three
dimensional. All possible critical points are found and their stability is
investigated. The phase portraits are plotted for both closed and open
models. The qualitative evolution of general solution at infinite values of
the phase space variables is investigated. The dilaton stabilization within
the framework of Damour-Polyakov mechanism is considered. Our results are
summarized in section 8.

\renewcommand{\theequation}{2.\arabic{equation}}

\setcounter{equation}{0}

\section{String effective action}

The beta functions for strings propagating in a background of massless
fields are the equations of motion of a certain master spacetime action
which can be computed as an expansion in the string tension $\alpha ^{\prime
}$. The general form of the corresponding $D$ - dimensional effective
action, including possible loop corrections, can be written as \cite
{lovelace}-\cite{gross}, \cite{green}, \cite{ekirit}, \cite{damour} 
\begin{equation}
S=\int d^Dx\sqrt{\left| \widetilde{G}\right| }\left[ -\widetilde{F}_R\left(
\varphi \right) \widetilde{R}-4\widetilde{F}_\varphi \left( \varphi \right)
\partial _M\varphi \widetilde{\partial }^M\varphi +\widetilde{L}_m\left(
\varphi ,\widetilde{G}_{MN},\psi \right) \right]  \label{actionst}
\end{equation}
where tilted letters specify the quantities in the string conformal frame, $%
\varphi $ - is the dilaton field, $\widetilde{R}$ denotes curvature scalar
of the $D$ - dimensional metric $\widetilde{G}_{MN}$, and $\psi $ stands for
all other degrees of freedom. Including the most important special cases,
the Lagrangian density $\widetilde{L}_m$ has the form 
\begin{eqnarray}
\widetilde{L}_m &=&-\sum_r\frac{\widetilde{F}_A\left( \varphi \right) }{%
2\left( \delta _r+1\right) !}F_r^2+\sum_i\widetilde{F}_{\chi _i}\left(
\varphi \right) \widetilde{G}^{MN}\partial _M\chi _i\partial _N\chi
_i-U\left( \varphi ,\chi _1,\chi _2...\right)  \label{matlag} \\
&&+\sum_i\left[ \widetilde{F}_{\psi _i}\left( \varphi \right) \stackrel{\_}{%
\widetilde{\psi }}_i\widetilde{\widehat{D}}\widetilde{\psi }_i-m_i\widetilde{%
F}_{m_i}\left( \varphi \right) \stackrel{\_}{\widetilde{\psi }}_i\widetilde{%
\psi }_i\right] +\cdots  \nonumber
\end{eqnarray}
Here $F_r=dA_r,r=1,2,...$ are the strengths of the $\delta _r$ - forms $A_r$
of both R-R and NS-NS types, $\chi _i$ are scalar fields (including various
moduli associated with compactification of extra dimensions), $\psi _i$
stand for fermion fields. The potential term $U$ includes, in particular, a
possible nonperturbative dilaton potential. Note that in the supergravity
theories deriving from superstring theories one naturally encounters form
fields of ranks 1-4. For example, the $10D$ type IIA string action contains
the universal Kalb-Ramond 2-form field $B_{MN}$ with strength $H_{MNP}$ in
the NS-NS sector, and 3-form field coming from $11D$ supergravity and second
rank field strength for the Kaluza-Klein vector $A_M$ in the R-R sector.

At the present stage of development of string theory we do not know the
global behaviour of the dilaton coupling functions $\widetilde{F}_K\left(
\varphi \right) $ beyond the fact that in the weak coupling limit $\left(
\varphi \rightarrow -\infty \right) $ they should admit an expansion 
\begin{equation}
\widetilde{F}_K\left( \varphi \right) =e^{-2\zeta \varphi }\left(
1+\sum_{l=1}^\infty Z_K^{(l)}e^{2l\varphi }\right)  \label{couplings}
\end{equation}
The first term in this expansion is the string tree-level contribution. At
this level the dilaton coupling is a uniform $e^{-2\varphi }$ ($\zeta =1$)
in the $NS-NS$ sector, and it does not couple (in string frame) to the $R-R$
sector field strengths ( $\zeta =0$ ). The dimensionless coefficient $%
Z_K^{(l)}$ represents the $l$ -loop contribution, and the parameter of the
loop expansion is $e^{2\varphi }$. Note that in the low-energy regime, with
broken supersymmetry there are no a priori obstacles to having couplings (%
\ref{couplings}) with $Z_K^{(l)}\neq 0$.

The effective action (\ref{actionst}) is written in the form directly
obtained from string $\sigma $ - model calculations (so-called string
frame). One may rewrite it in a different frame by making dilaton depended
rescaling field redefinition 
\begin{equation}
G_{MN}=\Omega ^{-2}(\varphi )\widetilde{G}_{MN}  \label{conftrans}
\end{equation}
By choosing the scale factor as

\begin{equation}
\Omega =\Omega _E,\quad \Omega _E\left( \varphi \right) =\widetilde{F}%
_R^{1/(1-n)}  \label{efactor}
\end{equation}
one obtains the Einstein (E-) frame action, where the graviton kinetic term
is diagonalized. The corresponding action takes the canonical form 
\begin{equation}
S=\int d^Dx\sqrt{\left| G\right| }\left[ -R-4F_\varphi \left( \varphi
\right) \partial _M\varphi \partial ^M\varphi +L_m\left( \varphi
,G_{MN},\psi \right) \right] ,  \label{actione}
\end{equation}
where the indices are now raised and lowered with $G_{MN}$. The function in
front of dilaton kinetic term and nongravitational Lagrangian density are
related to the corresponding string frame functions by 
\begin{equation}
4F_\varphi \left( \varphi \right) =-\frac n{n-1}\left( \frac{\widetilde{F}%
_R^{^{\prime }}}{\widetilde{F}_R}\right) ^2+4\frac{\widetilde{F}_\varphi }{%
\widetilde{F}_R},\qquad L_m=\Omega ^D\widetilde{L}_m\left( \varphi ,\Omega
^2G_{MN},\psi \right)  \label{effie}
\end{equation}
For $F_\varphi >0$ the dilaton field would be ghost-like, since its kinetic
term would be negative. For this reason we shall consider the case $%
F_\varphi <0$ . Under this assumption by introducing a new scalar field $%
\phi $ as 
\begin{equation}
d\phi =2\sqrt{-F_\varphi }d\varphi  \label{newphi}
\end{equation}
the dilaton kinetic term can be written in canonical form. At tree-level $%
F_\varphi =-1/(n-1)$ and the new scalar field is proportional to the
dilaton: 
\begin{equation}
\phi =\frac 2{\sqrt{n-1}}\varphi  \label{newphi1}
\end{equation}

In the general conformal frame related to the string one by transformation (%
\ref{conftrans}), the kinetic term of fermions also can be written in
canonical form rescaling the fermion fields (see also \cite{damour}) 
\begin{equation}
\psi _i=\left[ \widetilde{F}_{\psi _i}(\varphi )\Omega ^n(\varphi )\right]
^{1/2}\widetilde{\psi }_i  \label{fermrescale}
\end{equation}
The corresponding Lagrangian takes the form 
\begin{equation}
L_f=\sum_i\left[ \overline{\psi }_i\widehat{D}\psi _i-m_i\Omega (\varphi )%
\frac{\widetilde{F}_{m_i}\left( \varphi \right) }{\widetilde{F}_{\psi
_i}\left( \varphi \right) }\overline{\psi }_i\psi _i\right]  \label{fermlag}
\end{equation}
In the E-frame we have to substitute the conformal factor from (\ref{efactor}%
).

Another important conformal frame of scalar-tensor theories is the so-called
Jordan frame, in which the nongravitational part of the action does not
contain the scalar field. In this frame the laws of evolution of the
nongravitational fields take the same form as in multidimensional GR. For
example, in the case of fermion field $\psi _i$ , as it follows from (\ref
{fermlag}), the Jordan frame corresponds to the choice 
\begin{equation}
\Omega =\Omega _J^{(f)}=\widetilde{F}_{\psi _i}\left( \varphi \right) /%
\widetilde{F}_{m_i}\left( \varphi \right)  \label{fermjordan}
\end{equation}
This frame is very natural, because fermion matter couples directly only to
this metric, particles have constant masses and move on geodesics of its
metric. As we see from (\ref{fermjordan}) in the case of Damour-Polyakov
universality ansatz for the dilaton coupling functions, for fermions the
string and Jordan frames coincide.

For the action (\ref{actionst}) the Jordan frame, in general, can not be
realized. Let us consider the important special case, when in (\ref{actionst}%
) the dilaton dependence is factorized: 
\begin{equation}
\widetilde{L}_m\left( \varphi ,\widetilde{G}_{MN},\psi \right) =\widetilde{F}%
_L\left( \varphi \right) \widetilde{L}\left( \widetilde{G}_{MN},\psi \right)
.  \label{lag1}
\end{equation}
For these Lagrangians the Jordan frame exists, if the function $\widetilde{L}
$ has a certain conformal weight $\beta $: 
\begin{equation}
\widetilde{L}\left( \Omega ^2G_{MN},\psi \right) =\Omega ^{2\beta }%
\widetilde{L}\left( G_{MN},\psi \right)  \label{lag2}
\end{equation}
In general conformal frame the corresponding function is determined from the
last relation of (\ref{effie}) and has the form 
\begin{equation}
L_m=\Omega ^{D+2\beta }\widetilde{F}_L\left( \varphi \right) \widetilde{L}%
\left( G_{MN},\psi \right)  \label{lag3}
\end{equation}
Now the choice of the conformal factor in accordance with 
\begin{equation}
\Omega =\Omega _J,\quad \Omega _J^{D+2\beta }=\left| \widetilde{F}_L\left(
\varphi \right) \right| ^{-1}  \label{jordan1}
\end{equation}
leads to the Jordan frame with exception of the case $\beta =-D/2$ , when $%
\widetilde{L}_m$ is conformal invariant. For the $r$-forms considered above
the parameter $\beta =-r-1$ has the following values: $\beta =-3$ for Kalb -
Ramond field, $\beta =-2$ for gauge field. The scalar part of the Lagrangian
(\ref{matlag}) has a certain conformal weight ($\beta =-1$) if the potential
terms are absent and the condition (\ref{lag1}) is fulfilled if the
functions at kinetic terms are universal. Assuming the universality ansatz
for the dilaton coupling functions we see from (\ref{efactor}) and (\ref
{jordan1}) that the Jordan and E-frames coincide for massless scalar fields.
In particular this is the case at tree-level. Note that for the R-R fields
the string and Jordan frames coincide at tree-level.

\renewcommand{\theequation}{3.\arabic{equation}}

\setcounter{equation}{0}

\section{Cosmological model}

In this paper we shall consider homogeneous and isotropic cosmologies. The
corresponding E-frame metric has the form 
\begin{equation}
ds^2=N^2(t)dt^2-R^2(t)ds_n^2  \label{cosmetric}
\end{equation}
where $ds_n$ is the line element of a $n=D-1$ - dimensional space of
constant curvature, $N(t)$ and $R(t)$ are the lapse function and scale
factor. The choice of $N(t)$ corresponds to the various time coordinates.
For example, we have $N=1$ and $N=R(t)$ in the cases of synchronous and
conformal time coordinates, correspondingly. From the homogeneity of the
model it follows that the dilaton field should also depend on time only, $%
\varphi =\varphi (t)$. From the field equations we obtain that the
energy-momentum tensor corresponding to the metric (\ref{cosmetric}) is
diagonal and can be presented in the perfect fluid form 
\begin{equation}
T_M^N=diag(\varepsilon ,...,-p,...)  \label{emtensor}
\end{equation}
where $\varepsilon $ is the energy density and $p$ is the effective
pressure. If the corresponding nongravitational Lagrangian does not depend
on derivatives of the metric tensor, the values of these quantities in
various conformal frames are related by 
\begin{equation}
T_M^N=\Omega ^D\widetilde{T}_M^N  \label{transemt}
\end{equation}
with $\Omega =\Omega _{E,J}$ for the E- and Jordan frames respectively. The
functions $N$ and $R$ are frame dependent too: 
\begin{equation}
N(t)=\Omega ^{-1}(\varphi )\widetilde{N}(t),\qquad R(t)=\Omega ^{-1}(\varphi
)\widetilde{R}(t)  \label{transfactor}
\end{equation}

The E-frame evolution equations for the scale factor and dilaton field can
be written as 
\begin{eqnarray}
\stackrel{.}{H}+H\left( nH-\stackrel{.}{N}/N\right) &=&N^2b\varepsilon
-N^2k(n-1)/R^2  \label{equations} \\
\stackrel{..}{\phi }+\stackrel{.}{\phi }\left( nH-\stackrel{.}{N}/N\right)
&=&N^2\alpha \varepsilon /2  \nonumber \\
\stackrel{.}{\phi }^2-n(n-1)H^2 &=&-N^2\varepsilon +N^2kn(n-1)/R^2  \nonumber
\end{eqnarray}
where the overdots denote time derivatives, $k$ is the sign of the curvature
of the spatial sections ($k=-1,0,1$ for spatially-flat, closed and open
models respectively) and the following notation are introduced 
\begin{equation}
H=\stackrel{.}{R}/R,\qquad b=\frac{1-a}{2(n-1)}  \label{nothub}
\end{equation}
\begin{equation}
a=p/\varepsilon ,\qquad \alpha =\frac 1{\varepsilon \sqrt{\left| G\right| }}%
\frac{\delta L_m\sqrt{\left| G\right| }}{\delta \phi }  \label{nota}
\end{equation}
As is well known, the last equation of (\ref{equations}) is the constraint
equation, which is a consequence of first two ones and gives only a
restriction on the initial values. As we see from the second equation of (%
\ref{equations}) for the solutions with constant dilaton $\phi =\phi _0$ ,
the values $\phi _0$ have to be roots of the equation $\alpha (\phi )=0$.
These solutions coincide with those of GR. The existence of the function $%
\alpha (\phi )$ with appropriate zeroes provides a mechanism stabilizing the
dilaton. So far several schemes have been suggested to generate such a
function. The most popular among those is the generation of dilaton
potential due to non-perturbative effects, such as gaugino condensation in
the hidden sector. The corresponding cosmological equations follow from (\ref
{equations}) by choosing $\widetilde{L}_m=-\widetilde{V}(\varphi )$. Now the
E-frame functions $\varepsilon $ and $\alpha $ in RHS of equations (\ref
{equations}) are defined as 
\begin{equation}
\varepsilon =V(\varphi )\equiv \left[ \widetilde{F}_R(\varphi )\right] ^{%
\frac{1+n}{1-n}}\widetilde{V}(\varphi ),\quad \alpha \varepsilon =-V^{\prime
}(\varphi )  \label{potenergy}
\end{equation}
The corresponding qualitative analysis for spatially flat models have been
carried out in \cite{saharpot} without specifying the form of dilaton
potential.

An alternative mechanism stabilizing the dilaton at phenomenologically
acceptable values have been suggested in \cite{damour}. In this mechanism
(referred as Damour-Polyakov mechanism) the corresponding function $\alpha
(\phi )$ capable to stabilize the dilaton, is generated due to the
higher-loop corrections (see (\ref{couplings}) to the functions of dilaton
coupling with the matter sector. Below we shall consider just this case.

As a consequence of the dilaton dependence of Lagrangian $L_m$ the
energy-momentum tensor is acted upon by a dilaton gradient force: 
\begin{equation}
T_{M;N}^N=-\varepsilon \alpha \partial _M\phi  \label{conteq}
\end{equation}
In the cosmological context this equation takes the form 
\begin{equation}
\stackrel{.}{\varepsilon }/\varepsilon +n(1+a)H+\alpha \stackrel{.}{\phi }=0
\label{eqenergy}
\end{equation}
For the Jordan frame $\alpha =0$ and energy-momentum tensor is covariantly
conserved with respect to its metric.

Note that, in general, the quantities $a$ and $\alpha $ are functions of
time. The function $a$ is frame independent, and the values of the function $%
\alpha $ in the different conformal frames are related by 
\begin{equation}
\alpha _\Omega =\widetilde{\alpha }-(1-na)d\ln \Omega /d\phi ,  \label{alfa}
\end{equation}
where $\alpha _\Omega $ - corresponds to the conformal frame, defined by the
metric (\ref{conftrans}). In the Jordan frame the nongravitational
Lagrangian does not depend on dilaton field and therefore in this frame $%
\alpha =\alpha _J=0$. It follows from here that in the case of sources for
which the Jordan frame can be realized, in the string and E-frames we have
respectively 
\begin{equation}
\widetilde{\alpha }=\left( 1-na\right) d\ln \Omega _J/d\phi ,\qquad \alpha
=\left( 1-na\right) d\ln \left( \Omega _J/\Omega _E\right) /d\phi
\label{alfast}
\end{equation}
Note that the coefficient in these formulas, $1-na=T/\varepsilon $ is equal
to zero for the radiation. At tree-level the functions (\ref{alfast}) are
constant and in the E-frame 
\begin{equation}
\alpha =\frac{1-na}{\sqrt{n-1}}\left[ \frac{(n-1)\zeta }{n+2\beta +1}%
-1\right]  \label{alfatree}
\end{equation}
For NS-NS massless scalar fields ($\beta =-1,\zeta =1$) from (\ref{alfatree}%
) one has $\alpha =0$ and at this approximation Jordan and E-frames coincide.

For the field strength $F_r$ in (\ref{matlag}) there are two types of
ansatze which are compatible with the symmetries of the metric (\ref
{cosmetric}). For the first one, known as elementary, one has $r=n$ and 
\begin{equation}
F_{0M_1\cdots M_n}=const\cdot NR^nF_A^{-1}(\varphi )\epsilon _{M_1\cdots M_n}
\label{elementary}
\end{equation}
This form of antisymmetric tensor directly follows from the corresponding
field equations. In this case the Bianchi identity is vacuous. In $n=3$ such
an ansatz can be realized by 4-form field strength coming from 11D
supergravity. Note that for this field the additional Chern-Simons term is
contained in the Lagrangian (\ref{matlag}). But for the cosmological
backgrounds we consider this term is always zero.

The second ansatz is known as solitonic and is realized by forms with $r=n-1$
and 
\begin{equation}
F_{M_1\cdots M_n}=const\cdot \epsilon _{M_1\cdots M_n}  \label{solitonic}
\end{equation}
For the case $n=3$ as an example we can consider the Kalb-Ramond field. In (%
\ref{elementary}) and (\ref{solitonic}) the symbol $\epsilon _{M_1\cdots
M_n} $ takes the values 0 or 1 and is completely antisymmetric on all
indices. By calculating the corresponding energy-momentum tensors it can be
easily seen that in the case of solitonic ansatz the form field is
equivalent to the perfect fluid with equation of state $p=\varepsilon $ ($%
a=1 $), and the form field with elementary ansatz corresponds to the perfect
fluid with $p=-\varepsilon $ ($a=-1$). By taking into account that $\beta
=-n $ and $\beta =-(n+1)$ for solitonic and elementary ansatze, from (\ref
{alfatree}) one finds 
\begin{equation}
\alpha =\frac \epsilon {\sqrt{n-1}}[(n-1)\zeta +n-\epsilon ]
\label{alfaelsol}
\end{equation}
where $\epsilon =1,-1$ for the solitonic and elementary cases
correspondingly. Note that the tree-level cosmological backgrounds with
various types of NS-NS and R-R form fields were considered in \cite
{tseytlin0}, \cite{perry94}, \cite{copeland}-\cite{ricci}, \cite{lu}, \cite
{cop12}, \cite{lukas}, \cite{kaloper1}, \cite{kalop00},

In the paper \cite{saharcqg} for the general case of dilaton coupling
functions we have derived the cosmological solutions with arbitrary
curvature and dilaton, moduli and for the Kalb-Ramond fields. These
solutions were given in closed form in terms of integrations depending on
coupling functions. In present paper by using the qualitative methods we
shall investigate the main properties of cosmological models without the
specifying the dilaton coupling functions and for the general case of the
nongravitational source with barotropic equation of state. In particular,
the possibility of dilaton stabilization will be considered by using the
Damour-Polyakov mechanism.

\renewcommand{\theequation}{4.\arabic{equation}}

\setcounter{equation}{0}

\section{Dynamical system and special solutions}

In the following it will be convenient to work in the conformal gauge, when $%
N(t)=R(t)$. The corresponding time coordinate we will denote by index $c$.
By introducing a new independent variable $\tau $ and the function $h(\tau )$
in accordance with 
\begin{equation}
d\tau =\sqrt{H^2+k}dt_c,\qquad h=\frac{d\ln R}{d\tau }=\frac H{\sqrt{H^2+k}%
},\quad H=\frac{d\ln R}{dt_c}  \label{tau}
\end{equation}
(from the last equation of (\ref{equations}) it follows, that for the models
with $\varepsilon \geq 0$ the expression $H^2+k$ is nonnegative) and
substituting $N^2\varepsilon $ from the last equation into the first two
ones, the equations of motion (\ref{equations}) can be presented in the form
of third-order autonomous dynamical system 
\begin{eqnarray}
\frac{d\phi }{d\tau } &=&x,\qquad \frac{dx}{d\tau }=\left[ n(n-1)-x^2\right]
\left[ \alpha (\phi )/2-bhx\right] ,  \label{dynsis} \\
\frac{dh}{d\tau } &=&\left( 1-h^2\right) \left[ (n-1)(nb-1)-bx^2\right]  
\nonumber
\end{eqnarray}
This system is invariant under the transformations 
\begin{equation}
\tau \rightarrow -\tau ,\quad x\rightarrow -x,\quad h\rightarrow -h
\label{trans1}
\end{equation}
relating expanding and contracting models, and 
\begin{equation}
\phi \rightarrow -\phi ,\quad x\rightarrow -x,\quad \alpha \rightarrow
-\alpha   \label{trans2}
\end{equation}
relating the models, for two, in general different, functions $\alpha (\phi )
$ and $-\alpha (-\phi )$. For the source, satisfying the conditions (\ref
{lag2}) and (\ref{lag3}) with constant $a$ the function $\alpha $ is
determined by the relation (\ref{alfast}) and depends only on scalar field $%
\phi $. In the following we shall consider this case. The energy density can
be expressed via the solutions of the system (\ref{dynsis}) as 
\begin{equation}
R^2\varepsilon =\left( H^2+k\right) \left[ n(n-1)-x^2\right] .  \label{eps}
\end{equation}
It follows from here that the phase trajectories describing the models with
nonnegative energy density lie in region 
\begin{equation}
n(n-1)-x^2\geq 0.  \label{posenergy}
\end{equation}
If $t=\int Rdt_c$ denotes the E-frame comoving time, then for the
corresponding acceleration from the first of equations (\ref{equations}) one
obtains 
\begin{equation}
\frac{d^2R}{dt^2}=\frac{H^2+k}R\left[ (n-1)(nb-1)-bx^2\right] 
\label{acceleration}
\end{equation}
As it follows from here when 
\begin{equation}
a<2/n-1  \label{condinfl}
\end{equation}
one has accelerated expansion (in the E-frame), $d^2R/dt^2>0$ , in the
region 
\begin{equation}
bx^2<1-n(1+a)/2  \label{infregion}
\end{equation}
The segments of phase trajectories lying in this region describe an
inflationary type evolution.

In the condition (\ref{posenergy}) the sign of equality corresponds to the
pure gravi-dilaton models ($\varepsilon =0$). In this case the set of
cosmological equations is exactly integrable in terms of conformal time and
the corresponding solution have the form \cite{myers}, \cite{venez1}, \cite
{saharcqg} 
\begin{equation}
R=R_m\left| \sin \left[ \sqrt{k}(n-1)t_c\right] /\sqrt{k}\right| ^{1/(n-1)}
\label{gravidil1}
\end{equation}
\begin{equation}
\pm \ln \left| tg\left[ \sqrt{k}(n-1)t_c\right] /\sqrt{k}\right| =2\sqrt{%
\frac{n-1}n}\int \sqrt{-F_\varphi (\varphi )}d\varphi  \label{gravidil2}
\end{equation}
Given the dilaton coupling functions the second of these formula determine
the dilaton as a function of the conformal time. E-frame synchronous time
coordinate is defined by $t=\int Rdt_c$. This integral can be expressed
through the elementary functions for the spatially-flat models only and the
corresponding solution 
\begin{equation}
R\sim \left| t\right| ^{1/n},\quad \phi =const\pm \sqrt{1-1/n}\ln \left|
t\right|  \label{gravidile}
\end{equation}
lies in the basis of pre-big bang inflationary model proposed in \cite
{venez1}, \cite{gasperini3}. Note that for this solution the parameter $\tau 
$ is related to $t$ by 
\begin{equation}
\tau =\frac{sgnt}n\ln \left| t\right| .  \label{gravidiltau}
\end{equation}
The equation of pure gravi-dilaton phase trajectories of the dynamical
system (\ref{dynsis}) has the form 
\begin{equation}
x=\pm \sqrt{n(n-1)},\quad h=\mp \tanh ^k\left[ \sqrt{1-1/n}\left( \phi -\phi
_1\right) \right]  \label{gravidiltraj}
\end{equation}
with $\phi _1$ being an integration constant. These models represent an
invariant two-dimensional subspace of the three-dimensional phase space of (%
\ref{dynsis}). In figure 1 the corresponding phase portrait for the models
with increasing dilaton (upper sign) is plotted on the phase plane $(\phi
,h) $ compactified onto rectangle $(y,w)$ in accordance with 
\begin{equation}
e^\phi =\frac y{1-y},\quad 0\leq y\leq 1  \label{igrek}
\end{equation}
\begin{equation}
h=\frac w{1-\left| w\right| },\quad w=\frac H{\left| H\right| +\sqrt{H^2+k}%
},\quad -1\leq w\leq 1  \label{newzet}
\end{equation}
Note that $dw/dh=(1+\left| h\right| )^{-2}$ and the function $w=w(h)$ is
monotonic. The horizontal segments $w=\pm 1/2$ ($h=\pm 1$) correspond to the
E-frame expanding and contracting spatially-flat models. They split the
phase space into three invariant subspaces corresponding to the closed ($%
\left| w\right| <1/2$), expanding ($1/2<w<1$) and contracting ($-1<w<-1/2$)
open models. The segment $h=-1$ represents the pre-big bang inflationary
solution. At tree-level this solution in the string frame describes an
expansion of superinflationary type (pole inflation). As we see from figure
1 it is the late-time attractor for $k=\pm 1$ contracting (in the E-frame)
gravi-dilaton solutions. For these solutions inflation in the string frame
starts when trajectories approach the segment $h=-1$ sufficiently close and
they have finite time to inflate. To inflate sufficiently to solve horizon
and flatness problems of standard cosmology the solutions with $k=\pm 1$
have to approach the segment $h=-1$ at extremely early times. This
illustrates the fine tuning problem of the pre-big bang models considered in 
\cite{turner},\cite{Bousso}.

For the sources with the equation of state $p=\varepsilon $ (for example,
massless scalar field) one has $b=0$ and the first of the equations (\ref
{equations}) coincides with the corresponding vacuum equation. As a
consequence for these sources the E-frame scale factor is described by the
same expression (\ref{gravidil1}) (the corresponding solutions for the
dilaton see in \cite{saharcqg} ). Another important case when the dynamical
system (\ref{dynsis}) admits exact solution are the radiation-dominated
models.

\subsection{Radiation-dominated solution}

For the radiation $a=1/n$ and therefore according to (\ref{alfa}) $\alpha =0$%
. Integrating two last equations one has 
\begin{equation}
x^2=\frac{n(n-1)}{1+u},\quad h^2=1-\frac{ku_0u^2}{1+u},\quad u=\left( \frac
R{R_0}\right) ^{n-1},  \label{rad51}
\end{equation}
where $u_0$, $R_0$ are integration constants. As we see for the models with $%
k=1$ the possible values of the function $u$ are limited: $u\leq u_m$, where
the maximum value can be expressed via constant $u_0$ by means of relation $%
u_m=\left( 1+\sqrt{1+4u_0}\right) /2u_0$. The expressions for the Hubble
function and dilaton have the form 
\begin{equation}
\left| H\right| =\frac{\sqrt{1+u-ku_0u^2}}{\sqrt{u_0}u},\quad  \label{radhub}
\end{equation}
\begin{equation}
\phi =2\int \sqrt{-F_\varphi (\varphi )}d\varphi =const\pm \sqrt{\frac n{n-1}%
}\ln \left( 1+\frac 2u+2H\sqrt{u_0}\right) .  \label{raddilat}
\end{equation}
For the expanding models the upper (lower) sign in (\ref{raddilat})
corresponds to $\stackrel{.}{\phi }<0$ ($>0$). In the case of spatially flat
models ($k=0$) and $n=3$ the solution (\ref{raddilat}) for the scalar field
coincides with that given in \cite{nordvedt} for the scalar-tensor theories.
Note that as $u\rightarrow \infty $ (which is possible for open and flat
models only) $\phi $ move only a finite amount and $H\rightarrow 1$ for open
models and $H\rightarrow 0$ for spatially flat ones. The scale factor
dependence on the conformal time is determined from 
\begin{equation}
2u_0u=k+\sqrt{\left| 4u_0+k\right| }\cases{ \sin \left[ \sqrt{k}(n-1)\left(
t_c-t_0\right) \right] /\sqrt{k},\quad u_0>-k/4 \cr \cosh \left[ (n-1)\left(
t_c-t_0\right) \right] ,\quad k=-1,u_0<1/4 \cr \sqrt{u_0}\left[
(n-1)^2\left( t_c-t_0\right) ^2/(4u_0)-1\right] ,\quad k=0}  \label{radsol}
\end{equation}
(when $u_0=1/4$, $k=-1$ one has the solution $u=-2+2\exp \left[ (n-1)\left(
t_c-t_0\right) \right] $ ). The function $\phi (t_c)$ can be found either by
inserting (\ref{radsol}) into (\ref{raddilat}) or more simply by direct
integration of $\phi =const\cdot \int dt_c/u$ with the help of (\ref{radsol}%
). The last relation follows from the first integral of the second equation
of (\ref{equations}) in the case of radiation, $\stackrel{.}{\phi }%
R^{n-1}=const$. The E-frame synchronous time coordinate $t$ can be expressed
via $u$ as 
\begin{equation}
t=\frac{\sqrt{u_0}R_0}{n-1}\int \frac{u^{1/(n-1)}du}{\sqrt{1+u-ku_0u^2}}.
\label{radteu}
\end{equation}
Note that for closed models the segments $x=0$, $\left| h\right| \leq 1$ and
for open models the rays $x=0$, $\left| h\right| \leq 1$ are solutions of
the system (\ref{dynsis}) and represent the corresponding GR solutions. If
at the beginning of the radiation-dominated era $u=u_r$, then the total
variation of the dilaton field in this stage is equal to 
\begin{equation}
\phi _\infty -\phi _r=\pm \sqrt{\frac n{n-1}}\ln \frac{2+u_r+2\sqrt{%
1+u_r+u_0u_r^2}}{u_r\left( 1+2\sqrt{u_0}\right) }  \label{raddilvar}
\end{equation}
As it follows from here if, at the beginning of the radiation era the value
of $x$ is not too close to $\sqrt{n(n-1)}$, the total shift of the dilaton
is of order of one (see also \cite{nordvedt}).

The scale factor and comoving time coordinate for the corresponding
radiation-dominated solution in the string frame can be found by conformal
transformation and have the form 
\begin{equation}
\widetilde{R}=R_0\left[ u/\widetilde{F}_R(\varphi )\right] ^{1/(n-1)},\quad 
\widetilde{t}=\pm \frac{\sqrt{u_0}}{n-1}\int \frac{\widetilde{R}(u)du}{\sqrt{%
1+u-u_0u^2}}  \label{radstring}
\end{equation}
For the given coupling functions $\widetilde{F}_\varphi (\varphi )$ and $%
\widetilde{F}_R(\varphi )$ in the action (\ref{actionst}) we can obtain the
function $\varphi (u)$ by inverting the equation (\ref{raddilat}) and then
derive scale factor $\widetilde{R}(u)$ and comoving time $\widetilde{t}(u)$
from (\ref{radstring}). Therefore the formulae (\ref{raddilat}) and (\ref
{radstring}) determine the string frame radiation-dominated solution in
parametric form as a function of $u$, provided we specify the dilaton
coupling functions. By choosing $\widetilde{F}_\varphi =\widetilde{F}%
_R=e^{-2\varphi }$ one obtains the corresponding tree-level solutions.

\subsection{Special solutions for $\alpha =const$}

In the case $\alpha =const$ the dynamical system (\ref{dynsis}) admits two
classes of special solutions. They are important on account of that they are
early or late-time attractors for general solution, as we shall see below.
For the first of these solutions 
\begin{equation}
x=\alpha /2bh,\quad h=\pm 1  \label{specialsol1}
\end{equation}
It corresponds to the spatially flat models and has positive energy density
if $\left| \alpha /2b\right| <\sqrt{n(n-1)}$. The dependence of the scale
factor and scalar field on the E-frame comoving time is given by expressions 
\begin{equation}
R=const\cdot \left| t-t_0\right| ^{4b/\alpha _1},\quad \alpha _1=\alpha
^2+\frac n{n-1}(1-a^2)  \label{specialscale1}
\end{equation}
\[
\phi =const+\frac{2\alpha }{\alpha _1}\ln \left| t-t_0\right| 
\]
Here $t>t_0$ ($t<t_0$) corresponds to the expanding (contracting) models.
This solution describes an extended inflation when 
\begin{equation}
\alpha ^2/2b<2-n(1+a),  \label{extendinfl}
\end{equation}
the necessary condition of which is (\ref{condinfl}). The corresponding
energy density has the form 
\begin{equation}
\varepsilon =\frac 4{\alpha _1^2}\frac{\alpha _0^2-\alpha ^2}{(t-t_0)^2}%
,\quad \alpha _0=2b\sqrt{n(n-1)}  \label{sp1endens}
\end{equation}

The particular case of models with $\alpha =const$ are tree-level models,
for which the value of $\alpha $ is determined by (\ref{alfatree}). In this
case the corresponding string frame solution can be obtained by conformal
transformation 
\begin{equation}
\widetilde{R}=e^{\phi /\sqrt{n-1}}R,\quad \widetilde{t}=\int e^{\phi /\sqrt{%
n-1}}dt  \label{specialconf1}
\end{equation}
and has the form 
\begin{equation}
\widetilde{R}=const\cdot \left| \widetilde{t}-\widetilde{t}_0\right| ^{\frac{%
\alpha +\alpha _0/\sqrt{n}}{\alpha +\sqrt{n-1}\alpha _1/2}},
\label{specialst1}
\end{equation}
\[
\phi =const+\frac{\alpha \sqrt{n-1}}{\alpha +\sqrt{n-1}\alpha _1/2}\ln
\left| \widetilde{t}-\widetilde{t}_0\right| 
\]
when $\sqrt{n-1}\alpha \neq -1\pm \sqrt{1-n(1-a^2)}$, and the form 
\begin{equation}
\widetilde{R}=\widetilde{R}_0\exp \left[ \left( 1+2b\sqrt{n-1}/\alpha
\right) \widetilde{H}\widetilde{t}\right] ,\quad \varphi =\frac 12(n-1)%
\widetilde{H}(\widetilde{t}-\widetilde{t}_0)  \label{specialst11}
\end{equation}
otherwise (note that in the spatially-flat case the tree-level models are
exactly integrable in parametric form for general anisotropic barotropic
fluid (see \cite{gasperini3}, \cite{sah2})). Here $\widetilde{t}$ is the
string frame comoving time, related to the parameter $\tau $ and E-frame
comoving time by 
\begin{equation}
\tau /(4bh)=\alpha _1^{-1}\ln \left| t-t_0\right| =\left[ 2\alpha \sqrt{n-1}%
+\alpha _1\right] ^{-1}\ln \left| \widetilde{t}-\widetilde{t}_0\right| 
\label{comtimes1}
\end{equation}
For the considered special solution the relative contribution of dilaton to
the energy density is equal to $\alpha ^2/\alpha _0^2$. When tree-level
dilaton coupling to the nongravitational matter is trivial in the string
frame (the case $\zeta =0$ in (\ref{couplings})) from (\ref{specialst1}) one
obtains the solution previously considered in \cite{gasperini3}. The
analogous solutions in the case $\zeta =1$ are investigated in \cite{sah2}.
Note that for the case 
\begin{equation}
\alpha ^2=\frac{1-a}{n-1}[2-n(1+a)]  \label{specialcase}
\end{equation}
the scale factor is linear function on comoving time in both E- and string
frames. As it follows from (\ref{specialst1}) when 
\begin{equation}
\alpha =-\frac{1-a}{\sqrt{n-1}}  \label{conststscale}
\end{equation}
one has solution with constant string frame scale factor and logarithmic or
linear dilaton in the cases (\ref{specialst1}) and (\ref{specialst11})
respectively. For example, this condition is satisfied in the pure
gravi-dilaton model of noncritical strings, when in the action (\ref
{actionst}) $\widetilde{L}_m=\left( D-D_{cr}\right) e^{-2\varphi }/3$ , $%
D_{cr}$ - critical dimension ( =10 for superstrings). In this case $%
a=-1,\beta =0$ and in accordance with (\ref{alfatree}) $\alpha =-2/\sqrt{n-1}
$ , which coincides with (\ref{conststscale}). This solution corresponds to
the one of (\ref{specialst11}) with linear dilaton and was considered
previously in \cite{bachas}.

For the second class of special solutions of (\ref{dynsis}) with $\alpha
=const$ one has 
\begin{equation}
x=\pm \sqrt{(n-1)(n-1/b)},\quad h=\alpha /2bx  \label{specialsol2}
\end{equation}
The necessary condition for existence of this solution is (\ref{condinfl}).
In the E-frame the corresponding time dependence is determined by the
expressions 
\begin{equation}
R=H(t-t_0),\quad H=\pm \left| \frac{2-n(1+a)}{(n-1)\alpha ^2}(1-a)-1\right|
^{-1/2}  \label{specialscale2}
\end{equation}
\[
\phi =const+\alpha ^{-1}[2-n(1+a)]\ln R 
\]
with $t_0$ being integration constant. Here upper/lower sign corresponds to
the expanding/contracing models. At tree-level the same solution in the
string frame can be obtained by using the conformal transformation (\ref
{specialconf1}) and takes the form 
\begin{equation}
\widetilde{R}=\widetilde{H}(\widetilde{t}-\widetilde{t}_0),\quad \phi
=const+[1/\sqrt{n-1}+\alpha /(2-n(1+a))]^{-1}\ln \widetilde{R}
\label{specialst2}
\end{equation}
In the case 
\begin{equation}
\alpha =[n(1+a)-2]/\sqrt{n-1}  \label{constscalecl}
\end{equation}
one has a solution with constant string frame scale factor instead of (\ref
{specialst2}): 
\begin{equation}
\widetilde{R}=const,\quad \phi =\pm \sqrt{\frac 2{1+a}-n}\frac{\widetilde{t}-%
\widetilde{t}_0}{\widetilde{R}}  \label{curvsol1st1}
\end{equation}
In the limit $\left| h\right| \rightarrow 1$ in (\ref{specialsol2}) this
solution tends to the previous special solutions (\ref{specialsol1})
corresponding to the spatially flat models. Note that for the values $\alpha 
$ from (\ref{constscalecl}) one has $h_1^2=1-(n-1)(1+a)/(1-a)$, and hence
the solution (\ref{curvsol1st1}) corresponds to the closed models when $a>-1$
(recall that for the solution (\ref{specialsol2}) the condition (\ref
{condinfl}) must be satisfied). If in (\ref{constscalecl}) $a=-1$ ($\alpha
=-2/\sqrt{n-1}$) the solution (\ref{specialsol2}) corresponds to the
spatially-flat models ($h=\pm 1$) and has the form (\ref{specialst11}) with
the zero exponent. Note that the solution (\ref{specialsol2}) corresponds to
closed models when the condition (\ref{extendinfl}) is satisfied. As it have
been mentioned above in this case the solution (\ref{specialsol1}) describes
an extended inflation.

\renewcommand{\theequation}{5.\arabic{equation}}

\setcounter{equation}{0}

\section{Qualitative analysis of spatially-flat model}

For general case of the parameter $a$ the system (\ref{dynsis}) cannot be
solved explicitly. To obtain the generic features of the cosmological
evolution we shall use the dynamical systems methods. In this section we
shall focus on generic cosmological properties for the spatially-flat
models. For the cosmological model with flat space ($k=0$) from (\ref{tau})
one has $h=1,-1$ for the expanding and contracting models respectively. Now (%
\ref{dynsis}) is reduced to the second-order dynamical system 
\begin{eqnarray}
\frac{d\phi }{d\tau } &=&x,\quad \frac{dx}{d\tau }=\left[ n(n-1)-x^2\right]
\left( \alpha /2-bhx\right) ,  \label{flatsys} \\
\tau &=&const\cdot h\ln R,\quad h=\pm 1  \nonumber
\end{eqnarray}
Below we shall consider expanding models. The corresponding results for the
contracting ones can be obtained by transformation $\tau \rightarrow -\tau
,x\rightarrow -x$. The trajectories of the system (\ref{flatsys}),
describing the solutions with nonnegative energy density lie in the
horizontal strip (\ref{posenergy}) of the phase plane $(\phi ,x)$. The
boundary of this strip 
\begin{equation}
x=\pm \sqrt{n(n-1)}  \label{boundary1}
\end{equation}
is a solution of the system and corresponds to the pure gravi-dilaton
solutions (\ref{gravidil1}), (\ref{gravidil2}). At tree-level the function $%
\alpha $ is a constant (see (\ref{alfatree})) and the system (\ref{flatsys})
has another special solution (\ref{specialsol1}) with comoving time
dependence (\ref{specialscale1}) in the E-frame and (\ref{specialst1}) in
the string frame.

For the general tree-level spatially flat solutions the equation for the
phase trajectories have the form 
\begin{equation}
\left( \alpha -\frac{\alpha _0^2}\alpha \right) \phi =const+\ln \left[
\left( x-\frac \alpha {2b}\right) ^2\left( \sqrt{n(n-1)}+x\right) ^{\frac{%
\alpha _0}\alpha -1}\left( \sqrt{n(n-1)}-x\right) ^{-\frac{\alpha _0}\alpha
-1}\right]  \label{treetraj}
\end{equation}
where $\alpha _0$ is defined by (\ref{sp1endens}). As we see in the limiting
regions $\phi \rightarrow \pm \infty $ the general solution tends to one of
the special solutions (\ref{boundary1}), (\ref{specialsol1}).

For the case of radiation as a nongravitational source we have $a=1/n$ and
in accordance with (\ref{alfast}) $\alpha =0$. In this case the system (\ref
{flatsys}) has a solution with constant dilaton, which coincides with the
corresponding GR solution. The equation of the phase trajectories describing
the radiation-dominated solutions with variable dilaton has the form 
\begin{equation}
x=-\sqrt{n(n-1)}\tanh \left[ \alpha _0(\phi -\phi _1)/2\right] ,
\label{xrad}
\end{equation}
with $\phi _1$ being an integration constant. The corresponding $\tau $
dependence is given by the expression (the analogous solutions in
scalar-tensor theories of gravity for $n=3$ see \cite{nordvedt}) 
\begin{equation}
\phi =\phi _1\pm 2\sqrt{\frac n{n-1}}\ln \left[ e^{-(n-1)(\tau -\tau _0)/2}+%
\sqrt{1+e^{-(n-1)(\tau -\tau _0)}}\right]  \label{phirad}
\end{equation}
with a new integration constant $\tau _0$.

We now turn to the systematic analysis of the qualitative structure of the
phase portraits of dynamical system (\ref{flatsys}) for the general case of
the function $\alpha (\phi )$. The key aspect for this is to determine the
critical points and their stability. For the system (\ref{flatsys}) those
are the following points of the phase plane $(\phi ,x)$: 
\begin{equation}
(\phi _0,0),\quad \alpha (\phi _0)=0  \label{singpoints}
\end{equation}
Note that according to the relation (\ref{alfast}) the zeroes $\phi _0$ are
extremums of the function $\Omega _J/\Omega _E$. Here $\Omega _E$ have the
form (\ref{efactor}) and for the sources with the Lagrangians of given
conformal weight the function $\Omega _J$ is determined by (\ref{jordan1}).
In particular by assuming the Damour-Polyakov universality ansatz for the
higher genus dilaton couplings, one finds that the zeroes (\ref{singpoints})
are extremums of the dilaton coupling function.

Near the critical points the corresponding linear system coincides with the
equation of damped oscillator. The character of the critical points is
defined by the eigenvalues 
\begin{equation}
\lambda _{1,2}=-k_0\pm \sqrt{k_0^2+n(n-1)\alpha _0^{\prime }/2},\quad \alpha
_0^{\prime }\equiv \left( \frac{d\alpha }{d\phi }\right) _{\phi =\phi
_0},\quad k_0=\frac n4(1-a)  \label{eigvalflat}
\end{equation}
For the real $\lambda _i$ the angular coefficients of the separatrixes in
this point are equal to $k_i=\lambda _i$. The standard analysis (see, for
example, \cite{Bautin} ) leads to the following results.

\begin{enumerate}
\item  When $\alpha _0^{\prime }>0$ the eigenvalues $\lambda _i$ are real
and have different signs. The critical point is a saddle (see figure 2f).

In the case $\alpha _0^{\prime }<0$ the critical point (\ref{singpoints}) is
stable (asymptotically stable for $\alpha _0>0$) for the expanding models
and unstable for contracting ones. It is necessary to distinguish the
following cases.

\item  When $\alpha _0=0$ , $\alpha _0^{\prime }<0$ the critical point is a
center (see figures 3e and 3f). Near this point the solution for the scalar
field has the form 
\begin{equation}
\phi -\phi _0=\phi _1\cos \left( \omega _0\tau +\theta \right) ,\quad \omega
_0^2=-n(n-1)\alpha _0^{\prime }/2  \label{centvic}
\end{equation}
This case corresponds to sources with the equation of state $p=\varepsilon $%
, as it is the case for massless scalars and form-fields with solitonic
ansatz. For $D=4$ the antisymmetric Kalb-Ramond field is reduced to the
pseudoscalar axion field as well. The corresponding solutions are considered
in \cite{saharcqg}.

\item  $\alpha _0^{\prime }<-\alpha _0^2/8$. The critical point is a stable
sink (see figures 3a and 3d) and the solution in its vicinity represents
damped oscillations for the expanding models: 
\begin{equation}
\phi -\phi _0=\phi _1e^{-k_0\tau }\cos \left( \sqrt{\omega _0^2-k_0^2}\tau
+\theta \right)   \label{focvic}
\end{equation}

\item  $\alpha _0^{\prime }=-\alpha _0^2/8$. The eigenvalues are real and
equal to $-\sqrt{n(n-1)}\alpha _0/4$. The system is critically damped and
the corresponding point is an improper node (see figure 3b). Near of this
point the solution has the form 
\begin{equation}
\phi -\phi _0=(a_1+a_2\tau )e^{-k\tau }  \label{critvic}
\end{equation}
with $a_i$ being real constants.

\item  $-\alpha _0^2/8<\alpha _0^{\prime }<0$. The eigenvalues are real and
negative. The critical point is a stable node (figure 3c). As $\left|
\lambda _1\right| <\left| \lambda _2\right| $ near of this point all
trajectories except special ones, touch the eigenvector corresponding to the
eigenvalue $\lambda _1$ and the solution looks like 
\begin{equation}
\phi -\phi _0=b_1e^{\lambda _1\tau }+b_2e^{\lambda _2\tau }  \label{sadvic1}
\end{equation}

\item  And at last let us consider the degenerate case $\alpha _0^{\prime }=0
$, when $\lambda _1=0$. Let near of the critical point the function $\alpha
(\phi )$ has an expansion 
\begin{equation}
\alpha (\phi )=\frac 1{m!}\alpha _0^{(m)}\left( \phi -\phi _0\right)
^m+\cdots ,\quad m\geq 2.  \label{alfexp}
\end{equation}
The standard analysis (see, e.g. \cite{Bautin}) leads to the following
results:

\begin{enumerate}
\item  $m=2$. The critical point is an equilibrium state with a node sector
(stable for $h=1$, and unstable for $h=-1$ ) and with two saddle sectors.
The angular coefficients of the separatrixes are equal to $k_1=0,$ $k_2=-%
\sqrt{n(n-1)}\alpha _0/2$. In the case $\alpha _0^{\prime \prime }>0(<0)$
the trajectories of the node sector tend to the critical point from the left
(right) of the separatrix with coefficient $k_2$.

\item  $m=3$. The point has a character of a saddle (node) when 
\begin{equation}
1+4\alpha _0^{(3)}/3\alpha _0^4>0(<0)  \label{deg1}
\end{equation}

\item  $m\geq 4$. The critical point is a saddle.
\end{enumerate}
\end{enumerate}

Uniting the expressions (\ref{centvic}), (\ref{focvic}), (\ref{sadvic1}) in
the general formula 
\begin{equation}
\phi -\phi _0=\phi _1e^{a_1\tau }\cos (a_2\tau +a_3)  \label{genvic}
\end{equation}
with appropriately defined constants $a_i$, from the first of the equations (%
\ref{equations}) it is easy to find the behaviour of the Hubble function
near the critical point 
\begin{equation}
H=\frac 2{n(1+a)t}\left\{ 1+const\cdot e^{2a_1\tau }\left[ c_1+c_2\sin
(2a_2\tau +c_3)\right] \right\} ,\tau =\frac{\pm 2\ln \left| t\right| }{%
n(1+a)},\tau \rightarrow \infty  \label{hubvic}
\end{equation}
where upper/lower sign corresponds to the E-frame expanding/contracting
models, and constants can be expressed via $a_i$ and parameters $a,n$. The
first summand in the curly brackets corresponds to the GR solution and the
second one represents the small correction due to variable dilaton.

From the above analysis it follows that in the case of expanding models the
existence for the function $\alpha (\phi )$ of zero $\phi =\phi _0$ with $%
\alpha ^{\prime }(\phi _0)<0$ leads to an efficient mechanism for the
dilaton stabilization and relaxation of string effective gravity to the GR.
Indeed, for the relation of the dilaton variations at some time moments $t_i$
and $t_e$ we have 
\begin{equation}
\frac{\phi (t_e)-\phi _0}{\phi (t_i)-\phi _0}\sim \exp \left\{ -\left|
Re\lambda _1\right| \left[ \tau (t_e)-\tau (t_i)\right] \right\} =\left[ 
\frac{R(t_i)}{R(t_e)}\right] ^{\left| Re\lambda _1\right| }  \label{reldil}
\end{equation}
In the cases 3) and 4), when critical point is a sink or improper node one
has $\left| Re\lambda _1\right| =n(1-a)/4$. In the inflationary stage $a=-1$
and the relation of the scale factors at the beginning and at the end of
inflation is $\sim e^{-65}$ . If initially the dilaton shift $\sim 1$, one
receives that at the end of inflation the dilaton variations are strongly
suppressed ($n=3$)(see \cite{damour1}) 
\begin{equation}
\phi (t_e)-\phi _0<10^{-42}  \label{constdil}
\end{equation}
Assuming that for the various sources the zero of the function $\alpha (\phi
)$ is the same (in particular, this is the case when the dilaton couplings
are universal) the subsequent expansion stages (radiation- and matter-
dominated) leads to the further suppression of these variations: $\left|
\delta \phi \right| <10^{-49}$ at present epoch \cite{damour1}. Such
variations are much more far from observational restrictions following from
the tests of the equivalence principle. Thus this scheme leads to the
natural suppression of the dilaton variations in the expanding universe and
lies in the basis of dilaton stabilizing mechanism known as Damour-Polyakov
mechanism. Note that this scheme will work also in the case 5), if the value
of $\left| \alpha _0^{\prime }\right| $ is not too small.

In order to fully describe the qualitative evolution of the system (\ref
{flatsys}), we must investigate the behavior of the phase trajectories at
infinity of the phase plane. For this it is convenient to map the phase
plane onto a finite region. This can be done by introducing a new variable $%
y $ in accordance with (\ref{igrek}). In terms of this variable the system (%
\ref{flatsys}) takes the form 
\begin{equation}
\frac{dy}{d\tau }=y(1-y)x,\quad \frac{dx}{d\tau }=\left[ n(n-1)-x^2\right]
\left( \alpha (y)/2-bx\right) .  \label{flatsys1}
\end{equation}
The phase space $(y,x)$ of this dynamical system corresponding to the models
with nonnegative energy density is defined by the relations $0\leq y\leq 1$, 
$\left| x\right| \leq \sqrt{n(n-1)}$. The critical points $(y,x)$ on the
boundary of this region are the following points 
\begin{equation}
\left( j,\pm \sqrt{n(n-1)}\right) ,\quad j=0,1  \label{flatpoint1}
\end{equation}
\begin{equation}
\left( j,\alpha (j)/2b\right) ,\quad \left| \alpha (j)\right| \leq \alpha _0
\label{flatpoint2}
\end{equation}
The eigenvalues determining their stability are equal to 
\begin{equation}
\lambda _1=\pm (-1)^j\sqrt{n(n-1)},\quad \lambda _2=2bn(n-1)(1\mp \alpha
(j)/\alpha _0)  \label{eiginfflat}
\end{equation}
in the case of (\ref{flatpoint1}), and 
\begin{equation}
\lambda _1=(-1)^j\alpha (j)/2b,\quad \lambda _2=-bn(n-1)\left( 1-\alpha
^2(j)/\alpha _0^2\right)  \label{eiginfflat1}
\end{equation}
in the case of (\ref{flatpoint2}). From these expressions it follows that
the critical points are nodes or saddles. The points $(0,\sqrt{n(n-1)})$ and 
$(1,-\sqrt{n(n-1)})$ are unstable. Under condition $(-1)^j\alpha (j)<0$ the
point (\ref{flatpoint2}) is stable node and as $\phi \rightarrow \infty $
the trajectories tend to the solution described by the separatrix $x=\alpha
(j)/2b$ , having the form (\ref{specialscale1}) (by replacing $\alpha
\rightarrow \alpha (j)$) when $t\rightarrow \infty $. In the case $\alpha
(j)=0$ the eigenvalue $\lambda _1=0$ and the critical point (\ref{flatpoint2}%
) is degenerate. The additional analysis shows that if near $y=j,j=0,1$ one
has 
\begin{equation}
\alpha (y)=\frac 1{m!}\alpha ^{(m)}(j)(y-j)^m+\cdot \cdot \cdot ,\quad
\alpha ^{(i)}(j)=0,\quad i<m  \label{deginfinity}
\end{equation}
then for the expanding models the critical point (\ref{flatpoint2}) is an
stable node when $(1-2j)^{m+1}\alpha ^{(m)}(j)<0$ and a saddle point when $%
(1-2j)^{m+1}\alpha ^{(m)}(j)>0$.

For the sources with $a=1$ (for instance, massless scalars or axions) the
only critical points on the phase space boundary are the points (\ref
{flatpoint1}) with eigenvalues (\ref{eiginfflat}). In this case $b=0$ and $%
\lambda _2=\mp \sqrt{n(n-1)}\alpha (j)$. When $\alpha (j)=0$ all points of
the segment $y=j$, $\left| x\right| \leq \sqrt{n(n-1)}$ are critical points.

On the basis of the above analysis one can plot the phase portraits of the
dynamical system (\ref{flatsys}) for an arbitrary function $\alpha (\phi )$.
The corresponding phase-diagrams for the function $-\alpha (-\phi )$ can be
obtained by transformation (\ref{trans2}). Below in figures some
qualitatively different cases of the E-frame expanding models (upper sign in
(\ref{flatsys})) are presented. The portraits for the contracting models can
be obtained from here by using the transformation (\ref{trans1}). In figures
2 and 3 the phase plane $(\phi ,x)$ is mapped onto rectangle $0\leq y\leq 1$%
, $\left| x\right| \leq \sqrt{n(n-1)}$ according to (\ref{igrek}). The
horizontal segments $x=\pm \sqrt{n(n-1)}$ correspond to the pure
gravi-dilaton solutions (\ref{gravidil1}), (\ref{gravidil2}). For the point $%
(0,\pm \sqrt{n(n-1)})$ $((1,\pm \sqrt{n(n-1)}))$ one has $\tau =-\infty $ $%
(+\infty )$ and $\tau =+\infty $ $(-\infty )$ for the upper and lower signs
respectively. Therefore in accordance with (\ref{gravidiltau}) for the
corresponding values of the E-frame synchronous time coordinate one has $%
t=0(+\infty )$ and $t=+\infty (0)$.

The phase portrait of figure 2a corresponds to the tree-level models when $%
\alpha =const$ and it is assumed that $0<\alpha <\alpha _0$. Horizontal
segment $x=\alpha /2b$ represent the special solution (\ref{specialsol1})
considered in the subsection 4.2. In accordance with (\ref{comtimes1}) the
initial and final points of the evolution of this solution correspond to the
values $t=t_0$ and $t=+\infty $ of the E-frame synchronous time coordinate
and this solution is the late-time attractor for trajectories with $%
\varepsilon >0$. This special solution splits the phase space into two
subspaces. The trajectories corresponding to the first one evolve from the
weak coupling region ($\varphi \ll -1$) near the pure gravi-dilaton solution 
$x=\sqrt{n(n-1)}$ and finish the evolution in the strong coupling region.
For the second class of solutions (trajectories lying below of the segment $%
x=\alpha /2b$) the evolution starts in the strong coupling region at finite
time moment of the E-frame synchronous time near the gravi-dilaton solution
and finish that in the strong coupling region too at $t\rightarrow +\infty $%
. When $\alpha \rightarrow \alpha _0$ the special solution (\ref{specialsol1}%
) tends and in the limiting case coincides with the solution $x=\sqrt{n(n-1)}
$ . At further increase $\alpha >\alpha _0$ the qualitative character of the
phase portraits corresponding to the models with nonnegative energy density,
remains the same. Thus the phase portraits for $\alpha >\alpha _0$ can be
obtained from figure 2a by limitary transition $\alpha \rightarrow \alpha _0$
and are qualitatively equivalent to the phase portrait between the solutions 
$x=\alpha /2b$ and $x=-\sqrt{n(n-1)}$.

The tree-level phase portrait for the case $-\alpha _0<\alpha <0$ is plotted
in figure 2b. All solutions with $x>\alpha /2b$, except the pure
gravi-dilaton ones, finish their evolution in the weak coupling region near
the special solution (\ref{specialsol1}), which is the late-time attractor.
The solutions with $x<(>)\alpha /2b$ evolves from the strong (weak) coupling
region near the gravi-dilaton solution with decreasing (increasing) dilaton.
Note that now there are solutions for which during all evolution dilaton
remains in the weak coupling region. When $\alpha \rightarrow -\alpha _0$
the solution (\ref{specialsol1}) approaches to the gravi-dilaton solution $%
x=-\sqrt{n(n-1)}$. At further decrease $\alpha <-\alpha _0$ the qualitative
character of the phase portrait remains the same and is equivalent to that
of figure 2b between the solutions $x=\alpha /2b$ and $x=\sqrt{n(n-1)}$.
Note that in the considered cases the pure gravi-dilaton solutions are
unstable when $\left| \alpha \right| <\alpha _0$ and are stable when $\left|
\alpha \right| >\alpha _0$.

From (\ref{alfast}) it follows that for the radiation dominated models ($%
a=1/n$) one has $\alpha =0$ irrespective to the dilaton coupling functions.
The corresponding phase portrait of the dynamical system (\ref{flatsys}) is
presented in figure 2c. All points of the segment $0\leq y\leq 1$, $x=0$ are
critical and correspond to the solution with constant dilaton. For the
expanding models the solutions with variable dilaton originate from the
unstable nodes $(0,\sqrt{n(n-1)})$ and $(1,-\sqrt{n(n-1)})$ at some finite
time moment, and , except the pure gravi-dilaton solutions, tend to the
corresponding solutions with constant dilaton when $t\rightarrow +\infty $ .
The dilaton limitary value depends on the initial conditions. The equation
of the phase trajectories is determined by the relation (\ref{treetraj}),
and the dependence on $\tau $ - by relation (\ref{phirad}).

For the general case of dilaton coupling functions the character of the
phase diagrams of the dynamical system (\ref{flatsys}) depends on the number
and stability of the zeroes of $\alpha (\phi )$ and on the boundary values $%
\alpha (\pm \infty )$. When $\alpha (\pm \infty )>\alpha _0$ ($<-\alpha _0$%
), $\alpha (\phi )>0$ ($<0$) the only critical points are $(y=0,1;x=\pm 
\sqrt{n(n-1)})$ on the phase space boundary. For these cases the phase
portraits are qualitatively equivalent to the tree-level ones with $\alpha
=const>\alpha _0(<-\alpha _0)$ which are considered above. In figure 2d we
have plotted the case $\alpha (\pm \infty )<\alpha _0$, $\alpha (\phi )>0$.
The general solution with $\varepsilon >0$ approaches the special solution
with late-time asymptotic (\ref{specialscale1}), where one has to substitute 
$\alpha =\alpha (y=1)$. The case $\alpha (-\infty )>\alpha _0$, $\alpha
(+\infty )<\alpha _0$, $\alpha (\phi )>0$ is presented in figure 2e. Now all
models begin and finish their evolution in the strong coupling region and
the late-time behaviour is the same as in the previous case. When $-\alpha
_0<\alpha (-\infty )<0$ and $0<\alpha (+\infty )<\alpha _0$ the phase
portrait is depicted in figure 2f for the function $\alpha (\phi )$ with a
single zero. At this zero $\alpha ^{\prime }(\phi _0)>0$ and the
corresponding critical point is a saddle. The separatrixes of this saddle
point split the phase space into four qualitatively different regions. When $%
\alpha (-\infty )<-\alpha _0$ or $\alpha (+\infty )>\alpha _0$ the
corresponding portraits can be obtained from figure 2f by the corresponding
superposing of the points.

In all considered examples, apart the radiation dominated case and stable
separatrixes on figure 2f, the models finish their evolution on the phase
space boundary $\phi \rightarrow \pm \infty $ ( $y=0,1$ ). This models
cannot stabilize field $\phi $ at finite values. In figures 3a-3d we have
plotted the phase portraits for some simple models with variable $\alpha $ ,
leading to the dilaton stabilization as a result of cosmological evolution.
They are illustrations of Damour-Polyakov mechanism. The phase portrait in
figure 3a corresponds to the function $\alpha (\phi )$ with $\alpha (-\infty
)>\alpha _0$ , $\alpha (+\infty )<-\alpha _0$ and it is assumed that at the
zero $\phi =\phi _0$ of the function $\alpha (\phi )$ its derivative is
negative and satisfies to the condition $\left| \alpha ^{\prime }(\phi
_0)\right| >-\alpha _0^2/8$, when the critical point is a sink (case 3)).
For the assumed boundary values $\alpha (\pm \infty )$ all critical points
on the phase space boundary are saddle points. For all spatially-flat
models, apart the pure gravi-dilaton ones, the expansion of the universe
leads to the dilaton stabilization. When $0<\alpha (-\infty )<\alpha _0$, $%
\alpha (+\infty )<-\alpha _0$ the critical point appears on the boundary $%
y=0 $, having saddle nature and the unstable separatrix of which corresponds
to the special solution (\ref{specialsol1}) at $\phi \rightarrow -\infty $.
The phase portrait of this case is plotted in figure 3b, where it is assumed
the stable critical point is an improper node, $\left| \alpha _0\right|
=-\alpha _0^2/8$ (case 4) ). And at last when $0<\alpha (-\infty )<\alpha _0$%
, $-\alpha _0<\alpha (+\infty )<0$ the new critical saddle point appears on
the boundary $y=1$. The phase portrait is presented in figure 3c assuming
that $\alpha _0^{\prime }>-\alpha _0^2/8$ (case 5)), when the critical point
is a node. In all considered cases for all the models, except special ones,
corresponding to the separatrixes of critical points (\ref{flatpoint2}), the
initial stage of evolution is dilaton-dominated and the contribution of the
nongravitational matter can be neglected. The phase portraits for the other
possible combinations of the relations between $\alpha (\pm \infty )$ and $%
\alpha _0$ can be obtained from the cases considered above by transformation
(\ref{trans1}). In figures 2a-c are presented most favorable (from the point
of view of dilaton stabilization) cases, when for the all models with $%
\varepsilon >0$ the cosmological evolution transfers the system into the
states with small dilaton variations (all solutions tend to the models with
constant dilaton). As it follows from the expressions of the eigenvalues (%
\ref{eiginfflat}), (\ref{eiginfflat1}) at the phase space infinity there are
no stable points when 
\begin{equation}
\alpha (-\infty )\geq 0,\quad \alpha (+\infty )\leq 0  \label{condstabil}
\end{equation}
These conditions are necessary for that the all homogeneous and isotropic
models of effective string theory during the cosmological evolution have to
tend to the solutions with fixed dilaton. In the case of nongravitational
sources, for which the Jordan frame can be realized, the boundary value $%
\alpha (-\infty )$ is determined by the expression (\ref{alfatree}). For the
potential-dominated source $a=-1,\beta =0$ and hence in accordance with (\ref
{alfatree}) $\alpha (-\infty )=-2/\sqrt{n-1}$ and the point (\ref{flatpoint2}%
) with $j=0$ is stable node. As in the previous case for the part of
solutions the existence of the zeroes of the function $\alpha (\phi )$ with $%
\alpha _0^{\prime }<0$ does not lead to the relaxation toward the GR
solutions. In figure 3d we present the typical for these cases phase
portrait, when $-\alpha _0<\alpha (-\infty )<0$, the function $\alpha (\phi
) $ initially increases with increasing $\phi $, becomes zero with the
positive derivative in this point (left critical point in figure 3d, being
saddle point). Further the function reaches the maximum after which
decreases to the values $-\alpha _0<\alpha (+\infty )<0$. It is assumed that
for the second zero one has $\left| \alpha _0^{\prime }\right| >\alpha
_0^2/8 $ and it is a sink. The cosmological models leading to the final
fixed value of the dilaton begins from the unstable nodes $(0,\sqrt{n(n-1)})$%
, $(1,\sqrt{n(n-1)})$ in directions lying above the stable separatrixes of
the saddle point, corresponding to the left zero of the function $\alpha
(\phi )$. All the other models evolve toward the weak coupling region and
approach to the solution (\ref{specialsol1}) when $\tau \rightarrow +\infty $%
. The case $\alpha (-\infty )\leq -\alpha _0$ or $\alpha (+\infty )\leq
-\alpha _0$ can be obtained from figure 3d by simple superposing of the
points $(0,\alpha (-\infty )/2b)$ and $(0,-\sqrt{n(n-1)})$ or $(1,\alpha
(+\infty )/2b)$ and $(1,-\sqrt{n(n-1)})$ respectively.

For R-R fields when $1-na>0$ one has $\alpha (-\infty )<0$ and the point (%
\ref{flatpoint2}) with $j=0$ is stable node. Therefore there is a class of
the solutions with $\varphi \rightarrow -\infty $ when $t\rightarrow +\infty 
$ and for these solutions dilaton cannot be stabilized at finite values. For
massless $RR$ axion $\alpha (-\infty )=\sqrt{n-1}$, however in this case, as
it mentioned above, the zeroes of the function $\alpha (\phi )$ with $\alpha
_0^{\prime }<0$ are centers, near of which dilaton oscillates without
damping (case 2), and therefore these sources can not stabilize the dilaton.
This is the case for the sources with the equation of state $p=\varepsilon $
($a=1$) and for those $b=0$. For the function $\alpha (\phi )$ with a single
zero $\phi =\phi _0$, $\alpha ^{\prime }(\phi _0)<0$ the corresponding phase
portrait is similar to that of figure 3a with center instead of sink. In
figure 3e we have plotted the phase portrait for the more complicated
function with two zeroes: $\alpha ^{\prime }(\phi _0)>/<0$ for the
left/right one. For the corresponding limitary values we have $\alpha (\pm
\infty )<0$. There is a subclass of solutions with periodic dilaton. When $%
a=1$ and $\alpha (y=j)=0,j=0,1$ the all points of the segment $y=j$ are
critical. As an example in figure 3f we have plotted the case $\alpha
(y=0)>\alpha _0$, $\alpha (y=1)=0$. There are two classes of solutions: with
periodic dilaton, oscillating near the zero $\phi =\phi _0$, and models
which begin and finish their evolution in the strong coupling region.

On the basis of the analysis carried out above it is easy to investigate
more complicated cases of the function $\alpha (\phi )$. In the following
sections we shall consider the models with curved space.

\renewcommand{\theequation}{6.\arabic{equation}}

\setcounter{equation}{0}

\section{Spatially curved models with constant $\alpha $}

Another important case when (\ref{dynsis}) can be reduced to the second
order dynamical system are models with constant $\alpha $. In particular
they include tree-level models. Now the field $\phi $ is not contained into
the last two equations and we have the following second-order autonomous
dynamical system 
\begin{eqnarray}
\frac{dx}{d\tau } &=&\left[ n(n-1)-x^2\right] \left( \alpha /2-bhx\right)
\label{treecurv} \\
\frac{dh}{d\tau } &=&\left( 1-h^2\right) \left[ (n-1)(nb-1)-bx^2\right] , 
\nonumber
\end{eqnarray}
For the models with nonnegative energy density the phase space $\left(
x,h\right) $ of this system is determined by the condition $x^2\leq n(n-1)$,
and $h^2<1$ $(>1)$ for $k=1(-1)$.

\subsection{\protect\smallskip Critical points}

The critical points of (\ref{treecurv}) are the following:

\begin{enumerate}
\item  Points $(x,h)$ with coordinates 
\begin{equation}
\left( \pm \sqrt{n(n-1)},h_0\right) ,\quad h_0=-1,+1  \label{curvpont1}
\end{equation}
They represent the pure gravi-dilaton spatially-flat solutions (the
horizontal segments in figure 1). The corresponding eigenvalues and
eigenvectors are determined from the relations 
\begin{equation}
\lambda _2=\sqrt{n(n-1)}h_0\left( \alpha _0\mp h_0\alpha \right) ,\quad
\lambda _3=2h_0(n-1)  \label{curveignum1}
\end{equation}
\begin{equation}
\mathbf{n}_2=(1,0),\quad \mathbf{n}_3=(0,1)  \label{cueveigvec1}
\end{equation}
As it follows from here the points corresponding to the expanding models ( $%
h_0=1$) are always unstable and have a character of a saddle point when $%
\alpha _0\leq \pm \alpha $ and character of an unstable node when $\alpha
_0>\pm \alpha $. The directions of separatrixes are determined by the
eigenvectors (\ref{cueveigvec1}). In the case of node near of the critical
point all trajectories, except special ones, touch the special solution $h=1$
(eigenvector $\mathbf{n}_2$ ) when $\pm \alpha >\alpha _0-2\sqrt{1-1/n}$ and
the solutions $x=\pm \sqrt{n(n-1)}$ (eigenvector $\mathbf{n}_1$ ) in the
other cases. The behaviour of the trajectories near the points (\ref
{curvpont1}) with $h_0=-1$ is obtained by transformation (\ref{trans1}).

\item  Points 
\begin{equation}
(\pm \alpha /2b,\pm 1),\quad \left| \alpha \right| <\alpha _0
\label{curvpoint2}
\end{equation}
with the eigenvalues and eigenvectors 
\begin{equation}
\lambda _2=-n(n-1)bh_0\left( 1-\frac{\alpha ^2}{\alpha _0^2}\right) ,\quad
\lambda _3=-2n(n-1)bh_0\left( 1-\frac 1{nb}-\frac{\alpha ^2}{\alpha _0^2}%
\right)  \label{curveignum2}
\end{equation}
\begin{equation}
\mathbf{n}_2=(1,0),\quad \mathbf{n}_3\sim (1,2b/\alpha +4/n\alpha (\alpha
^2/\alpha _0^2-1)).  \label{curveigvec2}
\end{equation}
They correspond to the special solutions (\ref{specialsol1}). The point with
upper sign is a saddle when $\alpha ^2/\alpha _0^2>1-1/nb$, the direction of
the unstable separatrix of which is determined by vector $\mathbf{n}_3$, and
stable node in the other cases. The character of the point with lower sign
can be obtained by taking into account the transformations (\ref{trans1}).

\item  In the case $a<2/n-1$ one has also the following critical points 
\begin{equation}
\left( \pm \sqrt{(n-1)(n-1/b)},\pm h_1(\alpha )\right) ,\quad h_1(\alpha
)\equiv \frac \alpha {\alpha _0\sqrt{1-1/nb}}  \label{curvpoint3}
\end{equation}
with the eigenvalues and eigenvectors 
\begin{equation}
\lambda _i=\pm \frac{n-1}2h_1\left\{ -1\pm (-1)^i\left[
1-8(nb-1)(1-h_1^{-2})\right] ^{1/2}\right\} ,\quad i=2,3  \label{curveignum3}
\end{equation}
\begin{equation}
\mathbf{n}_{2,3}\sim (1,-2b\lambda _{3,2}/\alpha (n-1))  \label{curveigvec3}
\end{equation}
and corresponding to the special solutions (\ref{specialsol2}). Note that
for closed models $h^2<1$ and as it follows from (\ref{curveignum3}) the
eigenvalues have opposite signs and the point is a saddle. For the open
models $h^2>1$ and the point (\ref{curvpoint3}) with upper/lower sign and
for $\alpha >0$ is stable/unstable node (sink) when 
\begin{equation}
8(nb-1)(1-h_1^{-2})<1(>1)  \label{curvcond1}
\end{equation}
\end{enumerate}

In the E-frame the point (\ref{curvpoint3}) corresponds to the solution (\ref
{specialscale2}) considered in the subsection 4.2.

In the case 
\begin{equation}
\left| \alpha \right| =\alpha _0\sqrt{1-1/nb}  \label{alfaeq}
\end{equation}
in (\ref{curvpoint3}) we have $\left| h_1\right| =1$ and the critical point (%
\ref{curvpoint3}) corresponds to the spatially-flat models. Now this point
coincide with the point (\ref{curvpoint2}), and $\lambda _2=0$, that is it
is degenerate critical point. The standard qualitative analysis shows (see,
for example, \cite{Bautin}) that this point represents an equilibrium state
with one stable node sector and with two saddle sectors. The separatrix
between these sectors is the solution $x=\pm \sqrt{n(n-1)}$. For the node
sector $h^2>1$ and the corresponding solutions represent open models (for
this case the phase portrait of the open models see figure 4f ). The
trajectories of the saddle sectors correspond to the closed models, and the
phase portrait is qualitatively equivalent to the case of that figure 4e. In
the case (\ref{alfaeq}) the corresponding solution have the form 
\begin{eqnarray}
R &=&const\cdot \left| t\right| ,\quad \phi =const+\frac \alpha {2b}\ln
\left| t\right|  \label{alfaeqsol} \\
\widetilde{R} &=&const\cdot \left| \widetilde{t}\right| ,\quad \phi =const+%
\frac{\ln \left| \widetilde{t}\right| }{2b/\alpha +1/\sqrt{n-1}},\quad a\neq
-1  \nonumber
\end{eqnarray}
Then $a=-1$ we have the solution (\ref{specialst11}) with zero exponent.
Here the solution in the string frame is obtained at tree-level
approximation, when the dilaton field can be found from the relation $%
\varphi =\sqrt{n-1}\phi /2$.

In finite part of the phase plane the only critical points of the system (%
\ref{treecurv}) are points (\ref{curvpont1}), (\ref{curvpoint2}), (\ref
{curvpoint3}). For the closed models the phase space of the system (\ref
{treecurv}) is compact and there are no other critical points.

For the models with $k=-1$ in addition to the analysis of stability of the
critical points carried out above, it is necessary to investigate the
behaviour of phase trajectories at infinity of the phase space, when $%
h\rightarrow \pm \infty $. Due to the invariance with respect to the
transformation (\ref{trans1}) it is sufficient to consider the case $h>1$,
corresponding to the expanding models. To compactify the phase space let us
introduce a new variable $z=1/h$, $0<z<1$. In terms of independent variable $%
T$, $dT=hd\tau =d\ln R$ the dynamical system (\ref{treecurv}) takes the form 
\begin{eqnarray}
\frac{dx}{dT} &=&\left[ n(n-1)-x^2\right] (\alpha z/2-bx)  \label{opensysinf}
\\
\frac{dz}{dT} &=&z\left( 1-z^2\right) \left[ (n-1)(nb-1)-bx^2\right] 
\nonumber
\end{eqnarray}
the phase space of which is the rectangle $\left| x\right| \leq \sqrt{n(n-1)}
$, $0\leq z\leq 1$. Besides the points (\ref{curvpont1}), (\ref{curvpoint2})
and (\ref{curvpoint3}), the stability properties of which we have considered
above, there are the following critical points of the phase space $(x,z)$: 
\begin{equation}
\left( x=\pm \sqrt{n(n-1)},z=0\right)  \label{pointinf1}
\end{equation}
with the eigenvectors (\ref{cueveigvec1}) and eigenvalues 
\begin{equation}
\lambda _2=2bn(n-1),\quad \lambda _3=1-n  \label{pointeig1}
\end{equation}
and 
\begin{equation}
\left( x=0,z=0\right)  \label{pointinf2}
\end{equation}
with eigenvalues and eigenvectors 
\begin{equation}
\lambda _2=-bn(n-1),\quad \lambda _3=(n-1)(bn-1)  \label{pointeig2}
\end{equation}
\[
\mathbf{n}_2=(1,0),\quad \mathbf{n}_3\sim \left( 1,\frac{2(1/n-a)}{\alpha
(n-1)}\right) 
\]
The solutions corresponding to this point have the form 
\begin{equation}
R=t,\quad \phi =const  \label{flatspace}
\end{equation}
and for them $\varepsilon =0$. By making a coordinate transformation $%
t=(t^{\prime 2}-r^{\prime 2})^{1/2}$, $rt=r^{\prime }$ it is easy to see
that they correspond to the flat spacetime.

The critical points (\ref{pointinf1}) are saddles with stable separatrixes $%
x=\pm \sqrt{n(n-1)}$ and unstable separatrixes $z=0$. The point (\ref
{pointinf2}) is a stable node for $a>2/n-1$ and a saddle otherwise. For the
former case and expanding models in a neighborhood of the critical point (%
\ref{pointinf2}) the evolution is curvature dominated and the solution have
the form 
\begin{equation}
R\approx t\left\{ 1+\frac{c_0^2t^{2-n(1+a)}}{2[1-n(1-a)]}\right\} ,\quad
t\rightarrow +\infty  \label{asympfac6}
\end{equation}
\begin{equation}
\phi \approx const-\frac{c_0x_0}{n-1}t^{1-n}+\frac{\alpha c_0^2(n-1)}{%
2-n(1+a)}\frac{t^{2-n(1+a)}}{2(1/n-a)}  \label{asympdil6}
\end{equation}
with $c_0$ and $x_0$ being integration constants. Note that the asymptotic
formula (\ref{asympfac6}) is the same as in GR. When $a=1-1/n$ the solution
for the scale factor is $R\approx t+c_0^2\ln t/(4n-6)$. In the
radiation-dominated case the second time dependent term in the RHS of (\ref
{asympdil6}) is absent and $a=1/n$. In the case $a=2/n-1$ ($b=1/n$) the
points (\ref{curvpoint3}) and (\ref{pointinf2}) coincide and $\lambda _2=0$.
An additional analysis of the critical point (\ref{pointinf2}) shows that it
is a stable node.

\subsection{Qualitative evolution}

Having investigated all possible critical points and their stability we can
construct the phase-space diagrams for any values of parameters $\alpha $
and $a$. To plot phase portraits for closed and open models on the same
phase-space diagram it is convenient to use $w$ from (\ref{newzet}) as a
phase space variable instead of $z$ in (\ref{opensysinf}). In terms of $w$
the trajectories with nonnegative energy density lie in the rectangle 
\begin{equation}
\left| x\right| \leq \sqrt{n(n-1)},\quad \left| w\right| \leq 1
\label{curvpp}
\end{equation}
of the phase plane $(x,w)$. The boundary segments $x=\pm \sqrt{n(n-1)}$
correspond to the pure gravi-dilaton solutions, and $w=\pm 1/2$ correspond
to the spatially-flat models. The latter split the phase space into three
invariant subspaces corresponding to the closed, $\left| w\right| <1/2$,
expanding, $1/2<w<1$, and contracting, $-1<w<-1/2$, open models. The
dynamical system (\ref{opensysinf}) has a two dimensional parameter space $%
(a,\alpha )$, $\left| a\right| \leq 1$. In this space qualitatively
different regions are separated by curves 
\begin{equation}
a=2/n-1,\quad \alpha =\pm \alpha _0,\;\alpha =0,\quad \frac{\alpha ^2}{%
\alpha _0^2}=1-\frac{2(n-1)}{n(1-a)},  \label{parameterspace}
\end{equation}
where the latter corresponds to the condition $h_1^2=1$.

Below considering various cases of the parameters we shall assume $\alpha
\geq 0$. The phase portraits for negative $\alpha $ can be obtained by
taking into account the invariance of the system (\ref{opensysinf}) with
respect to $x\rightarrow -x$, $\alpha \rightarrow -\alpha $. Consider
qualitatively different cases in turn.

\begin{enumerate}
\item  For the values $a>2/n-1$, $\alpha >\alpha _0$ the only critical
points are those of (\ref{curvpont1}), (\ref{pointinf1}) and (\ref{pointinf2}%
). The corresponding phase portraits are presented in figure 4a. All
expanding models start at some finite moment of the E-frame comoving time
from the point $(x=-\sqrt{n(n-1)}$, $w=1/2)$. For the closed models they
terminate at a point $(x=-\sqrt{n(n-1)}$, $w=-1/2)$ at another finite time
moment. For the open models the trajectories approach to the flat spacetime
as $t\rightarrow +\infty $ with the asymptotic behaviour (\ref{asympfac6})
and (\ref{asympdil6}). The spatially-flat models with $w=1/2$ ($-1/2$) are
early (late)-time attractors for the open and closed models. All models are
noninflationary.

\item  When $a>2/n-1$, $0<\alpha <\alpha _0$, critical points are those of (%
\ref{curvpont1}), (\ref{curvpoint2}), (\ref{pointinf1}) and (\ref{pointinf2}%
). First two ones lie on the segments $w=\pm 1/2$ and correspond to the
spatially-flat models. Phase-space diagram of this case is plotted in figure
4b. As we see for the closed models there are three qualitatively different
classes of solutions separated by the separatrixes of the saddle points (\ref
{curvpoint2}): (i) models with decreasing dilaton ($x<0$), originating at
point $(-\sqrt{n(n-1)},1/2)$ in the finite past and approaching point $(-%
\sqrt{n(n-1)},-1/2)$ at the finite future, (ii) models with increasing
dilaton ($x>0$) originating at point $(\sqrt{n(n-1)},1/2)$ in the finite
past and approaching point $(\sqrt{n(n-1)},1/2)$ at the finite future, (iii)
models originating at point $(-\sqrt{n(n-1)},1/2)$ in the finite past and
approaching point $(\sqrt{n(n-1)},1/2)$ at the finite future. In addition,
two special solutions exist with semi-infinite lifetime, corresponding to
the separatrixes of the saddle (\ref{curvpoint2}). The first/second one
originate at point $(-\sqrt{n(n-1)},1/2)$ / $(\alpha /2b,1/2)$ in the
finite/infinite past and approach point $(-\alpha /2b,-1/2)$/ $(\sqrt{n(n-1)}%
,-1/2)$ in the infinite/finite future.

\item  The case $a=1/n$, $\alpha =0$ corresponds to the radiation-dominated
models. Phase-space diagram is plotted in figure 4c. It is symmetric under
reflection $x\rightarrow -x$. Vertical segments $x=0$ represent the GR
solutions for closed ($\left| w\right| <1/2$) and open ($\left| w\right|
>1/2 $) models. As we see only expanding GR solution is the attractor for
generic solution with varying dilaton. For all solutions dilaton field is
monotonic.

\item  When $a<2/n-1$, $\alpha >\alpha _0$ the critical points are (\ref
{curvpont1}), (\ref{curvpoint3}), (\ref{pointinf1}) and (\ref{pointinf2}).
The corresponding phase portrait is plotted in figure 4d. For points (\ref
{curvpoint3}) one has $h_1(\alpha )>1$ and it represents open models. We
have assumed that first of conditions (\ref{curvcond1}) takes place and this
point is a sink. In the case of upper sign it is the late-time attractor for
all expanding models with $\varepsilon >0$ and is described by expressions (%
\ref{specialscale2}), (\ref{specialst2}) in the E- and string frames. The
qualitative behavior of the closed models is similar to the case 1 with
difference that now exist the segments of trajectories with (\ref{infregion}%
) corresponding to the inflationary evolution in the E-frame. Note that the
closed models contain three stages: early-time dilaton-dominated,
intermediate inflationary and late-time dilaton dominated ones. The open
models also contain stages with inflationary evolution lying in the region (%
\ref{infregion}).

\item  In the case $a<2/n-1$, $\alpha <\alpha _0$, $h_1>1$ the new critical
points (\ref{curvpoint2}) appear on the segments $w=\pm 1/2$ corresponding
to the spatially-flat models. Phase-space diagram is plotted in figure 4e.
The qualitative behavior of the closed models is similar to the case 2, but
now an intermediate inflationary stages exists in the regions defined by (%
\ref{infregion}). For the open models there are two classes of solutions
separated by the unstable separatrixes of the saddles (\ref{curvpoint2}) and
(\ref{pointinf2}). In the case of expanding universe the models of both
classes ($w>1/2$) start near spatially-flat expanding solution in the finite
past with increasing and decreasing dilaton for the first and second classes
respectively. In addition there are two special solutions corresponding to
the unstable separatrixes of points (\ref{curvpoint2}) and (\ref{pointinf2}%
). The first one originates at point (\ref{curvpoint2}) in the finite past
and approaches point (\ref{curvpoint3}) with upper sign in infinite future.
The second solution starts at point (\ref{pointinf2}) corresponding to the
flat spacetime in finite paste and approaches point (\ref{curvpoint3}) in
infinite future.

\item  When $a<2/n-1$, $\alpha <\alpha _0$, $h_1<1$ the critical point (\ref
{curvpoint3}) corresponds to the closed models and is a saddle. The phase
portrait for this case is presented in figure 4f. There are two classes of
expanding open models ($w>1/2$) separated by unstable separatrix of point (%
\ref{pointinf2}). The point (\ref{curvpoint2}) is the late-time attractor
for both these classes. The trajectories of the first/second class start at
point $(-\sqrt{n(n-1)},1/2)$ / $(\sqrt{n(n-1)},1/2)$ in the finite past and
terminate at point (\ref{curvpoint2}) in the infinite future. After the
initial stage of dilaton-dominated expansion these models entry to the
inflationary stage. The separatrixes of the saddles (\ref{curvpoint3}) split
phase space of closed models ($\left| w\right| <1/2$) into seven invariant
subspaces. The trajectories above the stable separatrixes of the point (\ref
{curvpoint3}) originate at points $(\pm \sqrt{n(n-1)},1/2)$ in the finite
past and approach point (\ref{curvpoint2}) at the infinite future. Hence
these models have semi-infinite lifetime. The same is the case for the
closed models below the unstable separatrixes of the saddle (\ref{curvpoint3}%
) with lower sign. The qualitative behaviour of the other closed models is
the same as for corresponding ones in the cases 2 or 5. They have finite
lifetime.

\item  In the case $\alpha =\alpha _0\sqrt{1-1/nb}$ ($h_1=1$) critical
points (\ref{curvpoint2}) and (\ref{curvpoint3}) coincide and correspond to
the spatially-flat models. As it considered above the critical point is an
equilibrium state with one stable node sector corresponding to the open
models and with two saddle sectors corresponding to the closed models. The
corresponding phase-space diagram for the open/closed models is
qualitatively equivalent to that of figure 4 f/e.

\item  For the nongravitational source with $a=1$ one has $b=0$. In this
case all points of the segments $w=\pm 1$, $\left| x\right| \leq \sqrt{n(n-1)%
}$ are critical points. The corresponding phase-space diagram on the
rectangle $\left| x\right| \leq \sqrt{n(n-1)}$, $\left| w\right| \leq 1$
looks like of that in figure 1.
\end{enumerate}

\smallskip In the phase-space diagrams of figure 4 the point $(\sqrt{n(n-1)}%
,-1/2)$ corresponds to the pre-big bang gravi-dilaton models. As we see this
solution is late-time attractor for all contracting models in the cases of 1
and 4 (figures 4a and 4d) and for the part of models in the other cases of
figure 4.

We have considered tree-level phase portraits for $\alpha \geq 0$. The
corresponding phase space diagrams for $\alpha <0$ can be obtained by taking
into account the invariance of the dynamical system with respect to the
transformations $x\rightarrow -x$, $\alpha \rightarrow -\alpha $.

\renewcommand{\theequation}{7.\arabic{equation}}

\setcounter{equation}{0}

\section{ Qualitative analysis in general case}

Having discussed the cases of spatially-flat and tree-level curved models we
are now in a position to consider the general case of spatially curved
cosmological models without specifying dilaton couplings in the string
effective action. In the E-frame these models are described by the dynamical
system (\ref{dynsis}). The critical points of this system are the points of
the phase space $(\phi ,x,h)$ with coordinates 
\begin{equation}
\left( \phi _0,0,\pm 1\right) ,\quad \alpha (\phi _0)=0  \label{gencritpoint}
\end{equation}
where upper/lower sign corresponds to the expanding/contracting models. We
shall consider the first of these cases. The corresponding results for the
contracting models can be obtained by means of transformation (\ref{trans1}%
). Note that the points (\ref{gencritpoint}) are GR solutions for $k=0$. The
constant dilaton solutions for $k=1$ are presented by segments $\phi =\phi
_0 $, $x=0$, $\left| h\right| <1$, and for $k=-1$ by rays $\phi =\phi _0$, $%
x=0$, $\left| h\right| >1$.

Near the critical point (\ref{gencritpoint}) the equation for $h$ decouples
from the other two ones and the corresponding eigenvalues are equal to 
\begin{equation}
\lambda _{1,2}=-k_0\pm \sqrt{k_0^2+n(n-1)\alpha _0^{\prime }/2},\quad
\lambda _3=n(1+a)-2  \label{geneigval}
\end{equation}
The eigenvalues $\lambda _{1,2}$ define the character of the critical point
for the trajectories with $h=1$, corresponding to the spatially-flat models.
The different possible variants of this case have been considered in section
5. As it follows from (\ref{geneigval}) for the models $k=\pm 1$ the
critical point is unstable when $a>2/n-1$. The corresponding GR solutions
tend to the spatially-flat solution (point (\ref{gencritpoint}) with the
upper sign) as $t\rightarrow -\infty $. In the case $a<2/n-1$ the GR
solutions with $k=\pm 1$ tend to the solution (\ref{gencritpoint}) in the
limit $t\rightarrow +\infty $ (inflationary models). When additionally $%
\alpha _0^{\prime }<0$ (and therefore Re$\lambda _i<0$), the critical point
is stable and the solutions of string cosmology with variable dilaton tend
to the solution (\ref{gencritpoint}) too. Note that for point $(\phi
_0,0,-1) $ one has $\lambda _1>0$ and this point is always unstable.

Now we turn to the investigation of the behaviour of the trajectories at
infinity of the phase space. For this it is convenient to compactify the
phase space. In the case of the closed models it is sufficient to use the
mapping (\ref{igrek}). The corresponding dynamical system differs from (\ref
{dynsis}) by the first of the equations, which now has the form 
\begin{equation}
\frac{dy}{d\tau }=xy(1-y)  \label{igrekeq}
\end{equation}
and the phase space is the rectangular parallelepiped 
\begin{equation}
0\leq y\leq 1,\quad \left| x\right| \leq \sqrt{n(n-1)},\quad \left| h\right|
\leq 1  \label{genphasespace}
\end{equation}
The boundaries of this parallelepiped are invariant subspaces. The
trajectories on the boundaries $h=1,-1$ correspond to the spatially-flat
expanding and contracting models, respectively, and the trajectories on the
boundaries $x=\pm \sqrt{n(n-1)}$ represent the pure gravi-dilaton models.
For the latter case the equation of the phase trajectories have the form (%
\ref{gravidiltraj}).

For $\left| h\right| =1$ the critical points on the boundary of (\ref
{genphasespace}) are the points $(y,x,h)$ with coordinates ($\alpha =\alpha
(y)$) 
\begin{equation}
\left( j,\pm \sqrt{n(n-1)},h_0\right) ,\quad j=0,1,\quad h_0=1,-1
\label{gencritpoint2}
\end{equation}
\begin{equation}
\left( j,\alpha (j)/2b,h_0\right) ,\quad \left| \alpha (j)\right| \leq
\alpha _0  \label{gencritpoint3}
\end{equation}
When $h_0=1$ (the case $h_0=-1$ can be obtained by transformation (\ref
{trans1})) the eigenvalues $\lambda _{1,2}$ for these points are the same as
in the case of spatially flat models (formulas (\ref{eiginfflat}), (\ref
{eiginfflat1})), and 
\begin{equation}
\lambda _3=2(n-1)  \label{lamb31}
\end{equation}
for the point (\ref{gencritpoint2}), and 
\begin{equation}
\lambda _3=2n(n-1)b\left[ \alpha ^2(j)/\alpha _0^2+1/nb-1\right]
\label{lamb32}
\end{equation}
for the points (\ref{gencritpoint3}). Thus, for the models with $k=\pm 1$
points (\ref{gencritpoint2}) with $h_0=1$ are always unstable. Points $%
(j,\mp (-1)^j\sqrt{n(n-1)},\pm 1)$ are saddles always and correspond to an
infinite value of the E-frame comoving time. Points $(j,\pm (-1)^j\sqrt{%
n(n-1)},\pm 1)$ with upper/lower sign are unstable/stable nodes when $%
(-1)^j\alpha (j)<\alpha _0$ and saddles otherwise. They correspond to a
finite value of the comoving time coordinate and are singular.

The critical point (\ref{gencritpoint3}) with $h_0=1/-1$ is a
stable/unstable node when 
\begin{equation}
-\alpha _0\sqrt{1-1/nb}<(-1)^j\alpha (j)<0,\quad  \label{stabcond3}
\end{equation}
and a saddle otherwise. Note that the left of these conditions is possible
for $a<2/n-1$ only. For the expanding models point (\ref{gencritpoint3})
corresponds to a finite/infinite value of the comoving time when $%
(-1)^j\alpha (j)>/<0$. By taking into account (\ref{geneigval}) we conclude
that for $k=\pm 1$ all critical points with $h=1$ are unstable when $a>2/n-1$%
.

In the case $a<2/n-1$ there are critical points laying on the boundaries $%
y=0,1$ of parallelepiped (\ref{genphasespace}): 
\begin{equation}
\left( j,\pm \sqrt{(n-1)(n-1/b)},\pm h_1(\alpha (j))\right) ,\quad j=0,1
\label{gencritpoint4}
\end{equation}
where the function $h_1(\alpha )$ is defined by the relation (\ref
{curvpoint3}). These points correspond to the tree-level ones (\ref
{curvpoint3}). For the points (\ref{gencritpoint4}) the eigenvalues $\lambda
_{2,3}$ are determined by the relations (\ref{curveignum3}), and 
\begin{equation}
\lambda _1=\pm (-1)^j\sqrt{(n-1)(n-1/b)}  \label{lamb41}
\end{equation}
When $h_1^2<1$ the points (\ref{gencritpoint4}) correspond to the closed
models and are unstable, as the eigenvalues $\lambda _2$ and $\lambda _3$
have different signs. When $h_1^2>1$ the solutions corresponding to the
points (\ref{gencritpoint4}) describe open models. For the lower/upper sign
point (\ref{gencritpoint4}) with $j=0$ is stable/unstable source when $%
\alpha (y=0)<0$. In the case of $j=1$ and $\alpha (y=1)>0$ the point (\ref
{gencritpoint4}) is stable/unstable source for the upper/lower sign. For
other cases it is a saddle. Note that the stable sources correspond to the
expanding models with negative spatial curvature. In the case of
expanding/contracting models the value of the E-frame comoving time for the
point (\ref{gencritpoint4}) is $t=+\infty /-\infty $ when $\pm (-1)^j</>0$,
and this point corresponds to the finite time moment otherwise. For the
latter case the point (\ref{gencritpoint4}) presents a singularity for
cosmological solutions.

As for the open models $\left| h\right| >1$, to compactify phase space along
side with (\ref{igrek}) we shall to map 
\begin{equation}
z=1/h,\quad \left| z\right| \leq 1  \label{compact2}
\end{equation}
By introducing the independent variable $T$, $dT=hd\tau $, the corresponding
dynamical system can be written as 
\begin{eqnarray}
\frac{dy}{dT} &=&xyz(1-y),\quad \frac{dx}{dT}=\left[ n(n-1)-x^2\right]
\left[ \alpha z/2-bx\right]  \label{sysopen} \\
\frac{dz}{dT} &=&z\left( 1-z^2\right) \left[ (n-1)(nb-1)-bx^2\right] 
\nonumber
\end{eqnarray}
As in above we shall consider the expanding models (the case of contracting
models can be obtained by transformation $x\rightarrow -x$, $z\rightarrow -z$
). The phase space of the system (\ref{sysopen}) is the parallelepiped 
\begin{equation}
0\leq y\leq 1,\quad \left| x\right| \leq \sqrt{n(n-1)},\quad 0\leq z\leq 1
\label{spaceopen}
\end{equation}
and the critical points lie on the boundaries $z=1$ and $z=0$. For the first
of these cases ($h=1$) the critical points and their stability are the same
as in the previous case and correspond to the spatially-flat models. Thus we
shall consider the second case. At $z=0$ ($h=\infty $) the solutions of the
system (\ref{sysopen}) are the segments $z=0$, $y=constant$. All points $%
(y,x,z)$ of the segments 
\begin{equation}
(y,\,x=0,\,z=0),\quad 0\leq y\leq 1  \label{openinf1}
\end{equation}
and 
\begin{equation}
(y,\,x=\pm \sqrt{n(n-1)},\,z=0)  \label{openinf2}
\end{equation}
are critical. The eigenvalues corresponding to the axis $z$ are equal to 
\begin{equation}
\lambda _3=(n-1)(nb-1)  \label{lambinf3}
\end{equation}
for the points (\ref{openinf1}), and 
\begin{equation}
\lambda _3=1-n  \label{lambinf31}
\end{equation}
for the points (\ref{openinf2}). In the second case the critical points are
unstable. The points (\ref{openinf1}) are stable when $a>2/n-1$ and unstable
otherwise. By taking into account that for the first of these cases the
critical points for $z=1$ are unstable, we conclude that all models with $%
k=-1$ finish evolution on the segment $x=0$, $z=0$. Near the corresponding
critical point the trajectories of the dynamical system enter into this
point lying in the plane ($x,z$). In this plane the corresponding critical
point is an stable node with the eigenvalues 
\begin{equation}
\lambda _2=-nb(n-1),\quad \lambda _3=(n-1)(nb-1)  \label{eigopen}
\end{equation}
In the case $a<2/n-1$ the final point of the evolution for the models with $%
k=-1$ are the points on the boundary $z=1$, corresponding to the models with 
$k=0$, and the points (\ref{gencritpoint4}) under the conditions specified
above.

On the basis of the analysis carried out above it is possible to construct
the phase diagrams of the cosmological models for arbitrary function $\alpha
(\phi )$, defined by dilaton coupling functions in the string effective
Lagrangian. For given $a$ the position and character of the critical points
are determined by the zeroes of $\alpha (\phi )$ and by the boundary values $%
\alpha (\pm \infty )$. In the parameter space $(a,\alpha (y=j))$, $j=0,1$
the various qualitatively different regions are separated by curves (\ref
{parameterspace}) with $\alpha =\alpha (j)$. As in the tree-level case, to
plot the closed and open models on the same phase space diagram we shall use
the variable $w$ from (\ref{newzet}). In terms of the variables $(y,x,w)$
the phase space of the models with nonnegative energy density is the
parallelepiped 
\begin{equation}
0\leq y\leq 1,\quad \left| x\right| \leq \sqrt{n(n-1)},\quad \left| w\right|
\leq 1  \label{yxw}
\end{equation}
The trajectories lying on the sections $w=\pm 1/2$ represent invariant
subspaces and correspond to the spatially-flat models. For expanding
solutions the corresponding two-dimensional phase portraits are presented in
figures 2 and 3. The trajectories in the invariant subspaces $x=\pm \sqrt{%
n(n-1)}$ correspond to the pure gravi-dilaton solutions presented in figure
1 for increasing dilaton (upper sign). The phase portraits on the boundaries 
$y=0,1$ are the same as in the tree-level case with $\alpha =\alpha
(j)=const $ (see figure 4). The spatially-flat sections $w=\pm 1/2$ split
phase-space into three invariant subspaces corresponding to closed ($\left|
w\right| \leq 1/2$) and expanding ($w>1/2$) and contracting ($w<-1/2$) open
models. If $\phi =\phi _0$ is a zero of the function $\alpha (\phi )$ then
the dynamical system has solutions with constant dilaton. These solutions
are presented by the vertical segments $\left\{ \phi =\phi _0,x=0,\left|
w\right| \leq 1\right\} $ and correspond to the GR solutions. The horizontal
segment $x=\sqrt{n(n-1)}$, $w=-1/2$ presents spatially-flat pure
gravi-dilaton contracting solution with increasing dilaton. At tree-level
the corresponding string frame solution describes a superinflationary type
expansion and lies in the basis of pre-big bang inflationary model \cite
{venez1}, \cite{gasperini3}. As it follows from the above analysis this
solution is the late-time attractor for the spatially curved models with
additional sources when $\alpha (\phi =+\infty )>-\alpha _0$.

Let us consider two qualitatively different regions of the parameter $a$
separately.

\subsection{Qualitative evolution: $k=\pm 1,\quad a>2/n-1$}

In this case $nb<1$ and as it follows from ( \ref{dynsis}) the functions $%
h(\tau )$ and $w(\tau )$ are decreasing/increasing for the closed/open
models. There are no inflationary segments of trajectories. As possible
attractors one has the following critical points of the phase space $(y,x,w)$%
: 
\begin{equation}
(y,0,\pm 1)  \label{at1}
\end{equation}
\begin{equation}
(0,\mp \sqrt{n(n-1)},\mp 1/2),\quad \alpha (0)<\alpha _0  \label{at2}
\end{equation}
\begin{equation}
(1,\pm \sqrt{n(n-1)},\mp 1/2),\quad \alpha (1)>-\alpha _0  \label{at3}
\end{equation}
where upper/lower sign corresponds to the late/early-time attractors. The
other critical points are saddles and hence are unstable. The
contracting/expanding open models start/finish their evolution at the points
(\ref{at1}) with lower/upper sign, where $y$ is determined by the initial
conditions. The initial stage for the expanding models is dilaton dominated
and corresponding trajectories originate close to the pure gravi-dilaton
spatially-flat solutions. During the further expansion the open models
approach the point (\ref{at1}) with upper sign at the infinite future. In a
neighborhood of this point the solution have the asymptotic form (\ref
{asympfac6}), (\ref{asympdil6}) with $\alpha =\alpha (y)$ and $\phi $ tends
to a constant. Assuming the existence of the zeroes (\ref{gencritpoint})
with $\alpha ^{\prime }(\phi _0)<0$ for the closed models the field $\phi $
will initially evolve towards the value $\phi =\phi _0$. During the further
evolution $h$ becomes zero at some finite time moment and the universe
enters into the contracting phase. In this stage the friction term in the
scalar field equation changes sign taking away the scalar field from the
critical point. Hence in this case the GR solutions are not late/early-time
attractors for the general solution with varying dilaton.

Let us consider the qualitative evolution of the spatially curved models on
the phase-space $(y,x,w)$ for different boundary values $\alpha (j)$, $j=0,1$%
.

\begin{enumerate}
\item  $\alpha (j)>\alpha _0$, $j=0,1$. The critical points are possible
zeroes of the function $\alpha (\phi )$, (\ref{gencritpoint}), the points (%
\ref{openinf1}) and (\ref{openinf2}). All expanding models with $k=\pm 1$
and $\varepsilon >0$, apart from the possible trajectories with constant
dilaton, originate at point (\ref{at3}) with lower sign in the finite past
of the E-frame comoving time and approach point (\ref{at3}) with upper sign
in the finite future for the closed models, and point (\ref{at1}) with upper
sign in the infinite future for the open models. In the latter case the
limitary values of dilaton are determined by the initial conditions. The
pre-big bang gravi-dilaton solution $x=\sqrt{n(n-1)}$, $w=-1/2$ is the
attractor for all contracting models with $k=\pm 1$. The corresponding $3D$
phase-space diagram is plotted in figure 5a, assuming that the function $%
\alpha (\phi )$ has no zeroes. In this diagram the phase portraits on the
boundaries $y=0,1$ are similar to that of figure 4a.

\item  $\alpha (0)>\alpha _0$, $0<\alpha (1)<\alpha _0$. In addition to the
critical points of the previous case, the new unstable critical points (\ref
{gencritpoint3}) with $j=1$ appear for spatially-flat models. The
qualitative behaviour of the trajectories with $k=\pm 1$ ($w\neq \pm 1/2$)
is the same as in the case 1. When the function $\alpha (\phi )$ has no zero
the phase portrait on the section $w=1/2$ is similar to that of figure 2e.
Phase portrait on the boundary $y=1$ is equivalent to that of figure 4b.

\item  $\alpha (0)>\alpha _0$, $-\alpha _0<\alpha (1)<0$. Now the critical
point (\ref{gencritpoint3})\ is unstable for the spatially-flat models. The
qualitative picture for the spatially-curved models is the same as in the
case 1.

\item  $\alpha (0)>\alpha _0$, $\alpha (1)<-\alpha _0$. For the
spatially-flat models there are no stable critical points on the phase space
boundary. The boundary rectangles for the spatially-flat models are limitary
cycles for the spatially-curved solutions. All expanding models with $k=\pm
1 $ originate close to the rectangle with $w=1/2$ in the infinite past and
approach to the rectangle with $w=-1/2$ in the infinite future for the
closed models. The late/early-time behaviour of the expanding/contracting
open models is the same as in the previous cases. The corresponding phase
diagram is plotted in figure 5b. The phase portrait in the invariant
subspace $w=1/2$ is similar to that of figure 3a.

\item  $0<\alpha (0)<\alpha _0$, $\alpha (1)>\alpha _0$. The critical point (%
\ref{gencritpoint3}) with $j=0$ corresponds to the spatially-flat models and
is a saddle. The expanding open models originate at points (\ref{at2}) and (%
\ref{at3}) with lower signs in the finite past and approach point (\ref{at1}%
) at the infinite future. The corresponding limitary value of $y$ depends on
the initial conditions. In the simple case when $\alpha (\phi )>0$ for the
closed models there are three classes of solutions: (i/ii) models
originating at point (\ref{at3}) with lower sign in the weak coupling region
at the finite past and approaching point (\ref{at2})/(\ref{at3}) with upper
sign at the finite future, (iii) models originating at point (\ref{at2})
with lower sign in the finite past and approaching point (\ref{at3}) with
upper sign. Note that $(dx/d\tau )_{x=0}=n(n-1)\alpha (\phi )/2>0$ and
therefore the trajectories originating from (\ref{at2}) with lower sign can
not terminate at point (\ref{at2}) with upper sign. The early and late-time
asymptotics of these solutions are pure gravi-dilaton spatially-flat models.
For the trajectories of the class (ii) both these asymptotics lie in the
strong coupling region.

\item  $0<\alpha (j)<\alpha _0$, $j=0,1$. In addition to the points of the
previous case the new critical point, (\ref{at2}), $j=1$ appears. But this
point corresponds to the spatially-flat models and the qualitative features
of the spatially curved models remain the same as in the previous case.

\item  $0<\alpha (0)<\alpha _0$, $-\alpha _0<\alpha (1)<0$. The function $%
\alpha (\phi )$ has a zero with negative derivative. The corresponding
critical points are unstable. The qualitative behavior of the trajectories
for the spatially curved models with variable dilaton remains the same as in
the case 5. For the universe with positive curvature in addition to the
models of the case 5 there is also a class of solutions originating at point
(\ref{at2}) with lower sign in the finite past and approaching point (\ref
{at2}) with upper sign in the finite figure.. The corresponding phase
diagram is plotted in figure 6a.

\item  $0<\alpha (0)<\alpha _0$, $\alpha (1)<-\alpha _0$. For the models
with $k=\pm 1$ all critical points are saddles, except the points (\ref{at2}%
) and (\ref{at1}). The first of these points with lower/upper sign is
early/late-time attractor for expanding/contracting models. All closed
models originate in the weak coupling region in the finite past and
terminate in the same region at the finite future.

\item  Radiation-dominated models: $a=1/n$. The corresponding phase space
diagram is presented in figure 6b. The vertical segments $(y=const,x=0,w)$
are solutions of the dynamical system and correspond to the models with
constant dilaton. They split the phase-space into two invariant subspaces
with increasing ($x>0$) and decreasing ($x<0$) dilaton. The solutions of the
first/second class originate at point (\ref{at2})/(\ref{at3}) with lower
sign in the finite past and approach point (\ref{at1}) in the infinite
future for the open models and (\ref{at3})/(\ref{at2}) with upper sign in
the finite future for the closed models.
\end{enumerate}

\subsection{Qualitative evolution: $k=\pm 1,\quad a<2/n-1$}

Now there are segments of the trajectories with accelerated evolution lying
in the region (\ref{infregion}). When $\alpha _0^{\prime }<0$ the point (\ref
{gencritpoint}) with upper/lower sign is late/early-time attractor for the
models with varying dilaton. As in the case of the spatially-flat models
this provides an efficient mechanism for dilaton stabilization during the
expansion of the universe and is the basis of the Damour-Polyakov mechanism.
The other critical points lie on the boundary of the phase-space $(y,x,w)$.
In addition to the points (\ref{at2}) and (\ref{at3}) as possible attractors
at the phase-space infinity one has the following critical points 
\begin{equation}
(j,\,\pm \alpha (j)/2b,\,\pm 1/2),\quad -\alpha _0\sqrt{1-1/nb}<(-1)^j\alpha
(j)<0  \label{at4}
\end{equation}
\begin{equation}
\left( j,\,\mp (-1)^j\sqrt{(n-1)(n-1/b)},\,\frac{\mp (-1)^j\alpha (j)}{%
\alpha _0\sqrt{1-1/nb}}\right) ,\quad (-1)^j\alpha (j)<-\alpha _0\sqrt{1-1/nb%
}  \label{at5}
\end{equation}
where upper/lower signs correspond to the late/early-time attractors. Other
critical points (including (\ref{at1})) are saddles. The points (\ref{at5})
correspond to the open models. In a neighborhood of these points the
solution has the form (\ref{asympfac6}), (\ref{asympdil6}) with $\alpha
=\alpha (j)$. When $j=0$ dilaton is in the weak coupling region and the
corresponding asymptotics for string frame scale factor and dilaton are in
the form (\ref{specialst2}) or (\ref{curvsol1st1}). Let us consider
qualitatively different phase diagrams.

\begin{enumerate}
\item  $\alpha (j)>\alpha _0$, $j=0,1$. On the boundary of the
parallelepiped (\ref{yxw}) one has late/early-time attractors (\ref{at2})
and (\ref{at3}). The corresponding phase diagram is plotted in figure 7a
assuming that $\alpha (\phi )>0$. In this case there are no critical points
at the finite part of the phase-space. The expanding models originate at
point (\ref{at3}) with lower sign in the finite past and approach (i) point (%
\ref{at3}) with upper sign at the finite future for $k=1$ and (ii) point (%
\ref{at5}) with upper sign at the infinite future for $k=-1$. In the case of
models with positive spatial curvature $h$ becomes zero at some finite time
moment and then the universe enters into the contracting phase. Note that
when $a<2/n-1$ the function $h(\tau )$ is not monotonic and in general case
the expansion-contraction transition can take place several times. After the
initial dilaton-dominated decelerating expansion the models enter into the
inflationary stage which corresponds to $dh/dt>0$. For the closed models
after some amount of inflation the trajectories enter again into the
deflationary phase. After inflation the open models enter into the
oscillatory regime in a neighborhood of point (\ref{at5}) with upper sign.
In addition to these models there are also special solutions corresponding
to the unstable separatrixes of the points (\ref{at1}) with upper sign (see
figure 7a). In the case $\alpha (0)>\alpha _0$, $\sqrt{1-1/nb}<\alpha
(1)/\alpha _0<1$ the qualitative features of the phase diagram for the
spatially curved models is the same as in the figure 7a (note that this is
not the case for the spatially-flat models).

\item  When $\alpha (0)>\alpha _0$, $0<\alpha (1)<\alpha _0\sqrt{1-1/nb}$
the critical point (\ref{at5}) is a saddle and corresponds to the closed
models. Now at the infinity of the phase-space one has attractors (\ref{at3}%
) and (\ref{at4}) with $j=1$. The qualitative behaviour of the open models
is the same as in the previous case, with difference that all expanding
models approach point (\ref{at4}) with $j=1$. For the models with $k=1$ in
addition to the trajectories considered above there are models approaching
the point (\ref{at4}) with $j=1$ at the infinite future. These models
approach to the spatially-flat ones with the inflationary type final stage.
There are also special solutions corresponding to the stable separatrix of
saddle (\ref{at5}), $j=1$.

\item  $\alpha (0)>\alpha _0$, $-\sqrt{1-1/nb}<\alpha (1)/\alpha _0<0$. The
function $\alpha (\phi )$ has a zero with $\alpha ^{\prime }(\phi _0)<0$ and
the point (\ref{gencritpoint}) with upper sign is an attractor. Phase
diagram for this case is plotted in figure 7b. The segment $x=0$, $y=y(\phi
_0)$ presents the corresponding GR solutions. The expanding models with $%
k=-1 $ originate at point (\ref{at3}) with lower sign in the finite past and
approach point (\ref{gencritpoint}) with upper sign at infinite future.
There are also special solutions corresponding to the unstable separatrixes
of the saddles (\ref{at1}). For these models one has efficient dilaton
stabilization. In the case of the universe with positive spatial curvature
there are four classes of solutions having as early and late-time attractors
the points (i) (\ref{at3}) with lower sign and (\ref{at3}) with upper sign,
(ii) (\ref{at3}) with lower sign and (\ref{gencritpoint}) with upper sign,
(iii) point (\ref{gencritpoint}) with lower sign and (\ref{gencritpoint})
with upper sign, (iv) point (\ref{gencritpoint}) with lower sign and (\ref
{at3}) with upper sign. Dilaton stabilization by means of Damour-Polyakov
mechanism takes place for the models (iii) only. In the case of the models
(i) the initial and final stages of the cosmological evolution correspond to
the dilaton dominated decelerating expansion and contraction respectively,
whereas the intermediate stage is perfect fluid dominated with inflationary
type evolution. The models of (ii) and (iv) have semi-infinite lifetime. In
the case (iii) the universe originate in the infinite past at point (\ref
{gencritpoint}) with lower sign. After initial contraction the universe
enters into the expanding stage and approach point (\ref{gencritpoint}) with
upper sign as a result of inflation. For parameters region $\alpha
(0)>\alpha _0$, $-1<\alpha (1)/\alpha _0<-\sqrt{1-1/nb}$ the qualitative
features of the phase-space diagram for the spatially curved models is
similar to that of figure 7b.

\item  $\alpha (0)>\alpha _0$, $\alpha (1)<-\alpha _0$. There are no
attractors on the phase-space boundary. The boundary segments of the
rectangles $w=\pm 1/2$ are early (for upper sign) and late-time (for lower
sign) limitary cycles. The qualitative bahaviour of the models is the same
as in the previous case with difference that instead of nodes (\ref{at3})
now one has limitary cycles. The corresponding phase-diagram is plotted in
figure 8a, assuming a single zero for the function $\alpha (\phi )$.

\item  $\sqrt{1-1/nb}<\alpha (0)/\alpha _0<1$, $\alpha (1)>\alpha _0$. In
the case of positive function $\alpha (\phi )$ for the expanding open models
the only late-time attractor is point (\ref{at5}) with $j=1$. There are two
classes of such models originating at points (\ref{at2}) and (\ref{at3})
with lower signs in the finite past. In addition special solutions exist
corresponding to the unstable separatrixes of points (\ref{at1}) with upper
sign and (\ref{curvpoint3}) with $j=0$. The closed models can be divided
into three classes: originating at point (\ref{at3}) with lower sign and
approaching points (i) (\ref{at2}) and (ii) (\ref{at3}) with upper signs,
and trajectories originating at point (\ref{at2}) with lower sign and
approaching point (\ref{at3}) with upper sign. As in the case 5 for $a>2/n-1$
in the case of the positive function $\alpha (\phi )$ the trajectories
originating at point (\ref{at2}) with lower sign can not terminate at point (%
\ref{at2}) with upper sign.

\item  $\sqrt{1-1/nb}<\alpha (0)/\alpha _0<1$, $0<\alpha (1)<\alpha _0\sqrt{%
1-1/nb}$. As an attractors one has (\ref{at2}), (\ref{at3}) and (\ref{at4})
with $j=1$. The qualitative features for the open universe is similar to
those of the previous case with only difference that now instead of point (%
\ref{at5}), $j=1$ one has point (\ref{at4}), $j=1$ as late-time attractor
for the expanding models. In the case of the closed universe two new classes
of solutions appears (and the solutions obtained from these ones by time
inversion) in addition to previous case. They originate at points (\ref{at2}%
) and (\ref{at3}) with lower signs in the finite past and approach the point
(\ref{at4}), $j=1$ at the infinite future. The corresponding phase-space
diagram is depicted in figure 8b assuming a positive function $\alpha (\phi
) $.

\item  $\sqrt{1-1/nb}<\alpha (0)/\alpha _0<1$, $-\alpha _0\sqrt{1-1/nb}%
<\alpha (1)<0$. The function $\alpha (\phi )$ has a zero with negative
derivative. The phase-space diagram for the case of single zero is presented
in figure 9a. Critical points (\ref{gencritpoint}) are early (lower sign)
and late-time (upper sign) attractors. The other attractors are points (\ref
{at2}) and (\ref{at3}). The expanding open models originate at these points
with lower sign in the finite past and approach point (\ref{gencritpoint})
with upper sign at the infinite future. The evolution consists two stages:
initial decelerated dilaton-dominated and final inflationary type one. As a
result of the cosmological expansion the dilaton variations are strongly
suppressed. There are also special open solutions corresponding to the
unstable separatrixes of points (\ref{at1}) with upper sign. The closed
models originate at points (\ref{at2}), (\ref{at3}) and (\ref{gencritpoint})
with lower signs and approach these points with upper signs. As a
consequence one has nine qualitatively different types of the closed models.
There are also special solutions corresponding to the separatrixes of the
saddle point. For the models having the point (\ref{gencritpoint}) with
upper sign as a late-time attractor the dilaton is efficiently stabilized.
Final stage of these models present inflationary type expansion dominated by
barotropic perfect fluid with $a<2/n-1$. There are also models with
intermediate inflation. As we see from figure 9a, depending on the initial
conditions the both types of the inflationary evolution: standard and
pre-big bang, can be realized within the framework of the same model and
correspond to the trajectories having as a late-time attractor points (\ref
{gencritpoint}) and (\ref{at3}) with upper signs.

\item  $-\alpha _0\sqrt{1-1/nb}<\alpha (0)<0$, $\alpha (1)>\alpha _0$. The
function $\alpha (\phi )$ has a zero with $\alpha ^{\prime }(\phi _0)>0$.
The corresponding critical point is a saddle and the solutions with constant
dilaton are unstable. In figure 9b we have plotted phase space diagram
assuming a single zero for the function $\alpha (\phi )$. All attractors lie
on the phase-space boundary. These are points (\ref{at2}), (\ref{at3}), (\ref
{at4}) with $j=0$ and (\ref{at5}) with $j=1$. The latter point is a
late-time attractor for the expanding open models. These models can be
divided into three classes with qualitatively different behaviour
originating at points (\ref{at2}), (\ref{at3}) with lower signs and (\ref
{at4}) with $j=0$. All closed models except the solution with constant
dilaton originate at points (\ref{at2}), (\ref{at3}) with lower signs in the
finite past and approach the corresponding points with upper signs at the
finite future.

\item  In the case $-\alpha _0\sqrt{1-1/nb}<\alpha (0)<0$, $\sqrt{1-1/nb}%
<\alpha (1)/\alpha _0<1$ the phase-space diagram is similar to that of
previous case with difference that now one has point (\ref{at4}) with $j=1$
instead of (\ref{at5}) with $j=1$. The latter corresponds to the closed
models and is a saddle. For the closed universe in addition to the models of
previous case new class of solutions appears having as a late-time attractor
the point (\ref{at4}), $j=1$. This point correspond to the spatially-flat
models.
\end{enumerate}

\smallskip The phase diagrams for the another combinations of the values of $%
\alpha (0)$ and $\alpha (1)$ are qualitatively equivalent to the ones
considered above or can be obtained from those by taking into account the
invariance of the dynamical system under the transformation (\ref{trans2}).

\renewcommand{\theequation}{8.\arabic{equation}}

\setcounter{equation}{0}

\section{Conclusion}

We considered the evolution of a homogeneous and isotropic universe within
the framework of the string effective gravity with the string-loop
modifications of the dilaton couplings. (note that we have investigated the
most general case of coupling functions and our analysis is valid for any $D$%
-dimensional scalar-tensor theory. For the barotropic perfect fluid as a
nongravitational source the corresponding set of equations can be presented
in the form of the third-order autonomous dynamical system (\ref{dynsis}) by
using the variables (\ref{tau}). The model is specified by the barotropic
index $a$ and by the function $\alpha (\phi )$. The latter is determined by
dilaton couplings (see (\ref{alfast})) and is a constant at tree-level. As
an example of an additional source we consider form-fields, which naturally
arise in supergravities deriving from string theories.

For the pure gravi-dilaton, radiation-dominated and stiff fluid models the
solutions to the dynamical system (\ref{dynsis}) can be found in terms of
integrals containing combinations of the dilaton coupling functions. In the
case of constant $\alpha $ (which include tree-level models as a particular
case) there are two power-law special solutions, (\ref{specialsol1}) and (%
\ref{specialsol2}) with the E-frame time dependences (\ref{specialscale1})
and (\ref{specialscale2}), respectively. In addition to the pure
gravi-dilaton they are early or late-time attractors for the models with
varying $\alpha $. The first of these solutions corresponds to the
spatially-flat models and describes an extended inflation (in the E-frame)
in the case of (\ref{extendinfl}). In this case the second special solution
describes the universe with positive spatial curvature.

For the general case of $a$ and function $\alpha (\phi )$ to obtain the
generic features of the cosmological evolution we have used the dynamical
systems methods. We start with spatially-flat models in section 5 when
dynamical system is reduced to the second-order one. The only critical
points at the finite part of the phase space are those corresponding to the
zeroes of $\alpha (\phi )$ (for radiation $\alpha =0$ and all points of the
segment $x=0$ are critical). These points are saddles when $\alpha ^{\prime
}(\phi _0)\equiv \alpha _0^{\prime }>0$, centers for $\alpha _0^{\prime }<0$%
, $a=1$, and late-time attractors (for expanding universe) in otherwise. In
the last case we have an efficient dilaton stabilization as a result of the
cosmological expansion (Damour-Polyakov mechanism). Note that in the case of
the spatially-flat models this mechanism works for any $a<1$, despite the
fact that the stabilization efficiency depends on $a$. The most favorable
case is $a=-1$. The stability of the critical points at the phase space
infinity is investigated and various qualitatively phase portraits are
presented in figures 2 and 3. Another case when the dynamical system is
reduced to the second-order one are models with constant $\alpha $. The
phase space analysis of these models is given in section 6 and results are
presented in figure 4. Now dilaton can not be stabilized at finite values.

The general case of the spatially curved models with arbitrary function $%
\alpha (\phi )$ is considered in section 7. The corresponding phase space
diagrams are three-dimensional. Their qualitative structure are specified by
the zeroes of $\alpha (\phi )$, limitary values $\alpha (\phi =\pm \infty )$
and barotropic index $a$. The qualitatively different regions of the latter
are separated by $a=2/n-1$. The critical points corresponding to the zeroes
of $\alpha (\phi )$ with $\alpha _0^{\prime }<0$, $a<1$ are attractors for
the models with curved space when $a<2/n-1$ only. Another attractors tie at
the infinity of the phase space and correspond to the gravi-dilaton or
special solutions (\ref{specialsol1}) and (\ref{specialsol2}). For these
values of $a$ there are inflationary segments of trajectories lying in the
region (\ref{infregion}). For the models having the zeroes of $\alpha (\phi
) $ as a late-time attractor the final stage of the evolution is an
inflationary type expansion with efficient dilaton stabilization. As dilaton
variations in this stage are small it corresponds to the inflation in the
string frame as well. There is also a class of trajectories with pre-big
bang gravi-dilaton solution as a late-time asymptotic. Note that depending
on the initial conditions the both types of the inflationary evolution:
standard and pre-big bang, can be realized within the framework of the same
model. For various values of parameters the different scenarios of the
cosmological evolution are described in subsection 7.1 for $a>2/n-1$ and 7.2
for $a<2/n-1$. The corresponding phase space diagrams are presented in
figures 5 - 9.

In this paper we have considered the evolution of the cosmological models in
the E-frame. They have singularities at a finite E-frame comoving time. In
this higher curvature regime higher derivative terms due to string
corrections become important. The string frame solutions can be obtained by
transformations 
\begin{equation}
t_s=\int \Omega _E(\varphi )dt,\quad R_E=\Omega _E(\varphi )R
\label{Etostring}
\end{equation}
with frame comoving time $t_s$. By choosing dilaton coupling function one
could construct the metric in the string frame which is singularity free
over an infinite interval of the string comoving time. Analogous models for
the Jordan frame in the scalar-tensor theories were considered in \cite
{barrow1}, \cite{Rama}, where constraints are specified which lead to a
singularity free evolution in the Jordan frame. However as it have been
shown in \cite{kalop97} the conformal transformations alone cannot
completely remove the singularities from cosmological solutions in the
scalar-tensor models. In particular, gravitons follow geodesics in the
E-frame, which are incomplete and have a singularity at a finite E-frame
comoving time. Hence gravitational waves see the singularity.

\bigskip

\textbf{Acknowledgments}

\bigskip

This work is supported in part by grant no 96-855 of Ministry of Science and
Education of the Republic of Armenia.\newpage \

\pagebreak \newpage \newpage

{\large \textbf{Figure captions}}

\bigskip

\textbf{Figure 1.} Phase portrait for the gravi-dilaton models. The phase
space is mapped onto rectangle $(y,w)$, $0\leq y\leq 1$, $\left| w\right|
\leq 1$ according to (\ref{igrek}) and (\ref{newzet}). The horizontal
segments $w=\pm 1/2$ correspond to the spatially-flat models. The
trajectories lying in region $\left| w\right| >/<1/2$ describe the universe
with negative/positive spatial curvature.

\bigskip

\textbf{Figure 2.} Phase portraits for the spatially-flat models in the
phase space $(y,x)$, $0\leq y\leq 1$, $\left| x\right| \leq \sqrt{n(n-1}$.
(a) tree-level models with $0<\alpha <\alpha _0$, (b) tree-level models with 
$-\alpha _0<\alpha <0$, (c) radiation-dominated models $\alpha =0$, (d) $%
\alpha (\phi )>0$, $0<\alpha (\pm \infty )<\alpha _0$, (e) $\alpha (\phi )>0$%
, $\alpha (-\infty )>\alpha _0$, $0<\alpha (+\infty )<\alpha _0$, (f) $%
-\alpha _0<\alpha (-\infty )<0$, $0<\alpha (+\infty )<\alpha _0$ with a
single zero.

\bigskip

\textbf{Figure 3.} The same as in figure 2. (a) $\alpha (-\infty )>\alpha _0 
$, $\alpha (+\infty )<-\alpha _0$ with a single zero, (b) $0<\alpha (-\infty
)<\alpha _0$, $\alpha (+\infty )<-\alpha _0$ with a single zero as an
improper node, (c) $0<\alpha (-\infty )<\alpha _0$, $-\alpha _0<\alpha
(+\infty )<0$, with a single zero as a node, (d) $-\alpha _0<\alpha (\pm
\infty )<0$ with two zeroes, (e) stiff fluid ($a=1$), $\alpha (\pm \infty
)<0 $ with two zeroes, (f) stiff fluid, $\alpha (-\infty )>\alpha _0$, $%
\alpha (+\infty )=0$.

\bigskip

\textbf{Figure 4.} Phase portraits for the models with constant $\alpha $
compactified onto rectangle $(x,w)$, $\left| x\right| \leq \sqrt{n(n-1}$, $%
\left| w\right| \leq 1$. Vertical segments $x=\pm \sqrt{n(n-1)}$ correspond
to the gravi-dilaton models. The trajectories lying in region $\left|
w\right| >/<1/2$ describe the universe with negative/positive spatial
curvature. (a) $a>2/n-1$, $\alpha >\alpha _0$, (b) $a>2/n-1$, $0<\alpha
<\alpha _0$, (c) radiation-dominated models, $a=1/n$, $\alpha =0$, (d) $%
a<2/n-1$, $\alpha >\alpha _0$, (e) $a<2/n-1$, $\alpha <\alpha _0$, $h_1>1$,
(f) $a<2/n-1$, $\alpha <\alpha _0$, $h_1<1$.

\bigskip

\textbf{Figure 5.} Phase space diagrams for the spatially curved models with 
$a>2/n-1$ compactified onto parallelepiped $(y,x,w)$, $0\leq y\leq 1$, $%
\left| x\right| \leq \sqrt{n(n-1}$, $\left| w\right| \leq 1$. Sections $%
w=\pm 1/2$ correspond to the spatially-flat models. The trajectories lying
in region $\left| w\right| >/<1/2$ describe the universe with
negative/positive spatial curvature. (a) $\alpha (y=0,1)>\alpha _0$, $\alpha
(y)>0$, (b) $\alpha (0)>\alpha _0$, $\alpha (1)<-\alpha _0$.

\bigskip

\textbf{Figure 6.} The same as in figure 6. (a) $0<\alpha (0)<\alpha _0$, $%
-\alpha _0<\alpha (1)<0$, (b) radiation-dominated models, $a=1/n$, $\alpha
=0 $.

\bigskip

\textbf{Figure 7.} The same as in figure 5 for models with $a<2/n-1$. (a) $%
\alpha (y=0,1)>\alpha _0$, $\alpha (y)>0$, (b) $\alpha (0)>\alpha _0$, $-%
\sqrt{1-1/nb}<\alpha (1)/\alpha _0<0$.

\bigskip

\textbf{Figure 8.} The same as in figure 7. (a) $\alpha (0)>\alpha _0$, $%
\alpha (1)<-\alpha _0$, with a single zero for $\alpha (y)$, (b) $\sqrt{%
1-1/nb}<\alpha (0)/\alpha _0<1$, $\alpha (1)>\alpha _0$, with positive
function $\alpha (y)$.

\bigskip

\textbf{Figure 9.} The same as in figure 7. (a) $\sqrt{1-1/nb}<\alpha
(0)/\alpha _0<1$, $-\sqrt{1-1/nb}<\alpha (1)/\alpha _0<0$, with a single
zero for $\alpha (y)$, (b) $-\sqrt{1-1/nb}<\alpha (0)/\alpha _0<0$, $\alpha
(1)>\alpha _0$, with a single zero for $\alpha (y)$.

\end{document}